\documentclass[12pt]{article}
\usepackage[latin9]{inputenc}
\usepackage[a4paper]{geometry}
\usepackage{float}
\usepackage{mathtools}
\usepackage{amsmath}
\usepackage{amsthm}
\usepackage{amssymb}
\usepackage{esint}
\usepackage{pifont}
\usepackage[numbers,sort&compress]{natbib}
\usepackage[unicode=true,
 bookmarks=true,bookmarksnumbered=false,bookmarksopen=false,
 breaklinks=false,pdfborder={0 0 0},pdfborderstyle={},backref=false,colorlinks=false]
 {hyperref}

\makeatletter
\numberwithin{equation}{section}
\numberwithin{figure}{section}
\theoremstyle{plain}

\theoremstyle{definition}

\theoremstyle{plain}

\theoremstyle{plain}

\theoremstyle{plain}

\usepackage{braket}
\usepackage{feynmf}
\usepackage{subfigure}

\makeatother

\providecommand{\corollaryname}{Corollary}
\providecommand{\definitionname}{Definition}
\providecommand{\lemmaname}{Lemma}
\providecommand{\propositionname}{Proposition}
\providecommand{\theoremname}{Theorem}

\begin{document}


\begin{titlepage}

\thispagestyle{empty}

\begin{flushright}
\end{flushright}

\vspace{.4cm}
\begin{center}
\noindent{\large \bf Reflected Entropy for an Evaporating Black Hole}\\
\vspace{2cm}

Tianyi Li, Jinwei Chu, Yang Zhou
\vspace{1cm}

{\it
Department of Physics and Center for Field Theory and Particle Physics, \\
Fudan University, Shanghai 200433, China\\
}

\end{center}

\vspace{.5cm}
\begin{abstract}
We study reflected entropy as a correlation measure in black hole evaporation. As a measure for bipartite mixed states, reflected entropy can be computed between black hole and radiation, radiation and radiation. We compute reflected entropy curves in three different models: 3-side wormhole model, End-of-the-World (EOW) brane model in three dimensions and two-dimensional eternal black hole plus CFT model. For 3-side wormhole model, we find that reflected entropy is dual to island cross sections. The reflected entropy between radiation and black hole increases at early time and then decreases to zero, similar to Page curve, but with a later transition time. The reflected entropy between radiation and radiation first increases and then saturates. For the EOW brane model, similar behaviors of reflected entropy are found.

We propose a quantum extremal surface for reflected entropy, which we call quantum extremal cross section. In the eternal black hole plus CFT model, we find a generalized formula for reflected entropy with island cross section as its area term by considering the right half as the canonical purification of the left. Interestingly, the reflected entropy curve between the left black hole and the left radiation is nothing but the Page curve. We also find that reflected entropy between the left black hole and the right black hole decreases and goes to zero at late time. The reflected entropy between radiation and radiation increases at early time and saturates at late time.
\end{abstract}

\end{titlepage}

\setcounter{tocdepth}{3}
{\hypersetup{linkcolor=black}\tableofcontents}

\newpage

\section{Introduction}

The black hole evaporation process is expected to be unitary, therefore the Von Neumann entropy of the outgoing radiation should follow Page curve~\cite{Page:1993wv,Page:2013dx,Hawking:1976ra}, in which the entropy first increases and then decreases at so-called Page time. The change happens because, treated as entanglement entropy between the radiation and the black hole microstates, the entropy of the radiation can not exceed the remaining black hole entropy. In recent breakthrough works, a Page curve was computed in asymptotically AdS black hole plus conformal field theory reservoir~\cite{Penington:2019npb,Almheiri:2019psf}. In particular Page curve has been reproduced explicitly in Jackiw-Teitelboim (JT) gravity in AdS$_2$ without assuming unitarity~\cite{Almheiri:2019hni}. The key step to reproduce Page curve is to employ the island formula for the Von Neumann entropy of radiation, which was inspired from the quantum extremal surface formula for holographic entanglement entropy~\cite{RT:RT-formula,HRT:HRT-formula,Faulkner:2013ana,Engelhardt:2014gca}. Further justifications~\cite{Almheiri:2019qdq,Penington:2019kki} and generalizations have been explored~\cite{Rozali:2019day,Brown:2019rox,Chen:2019iro,Akers:2019lzs,Pollack:2020gfa,Liu:2020gnp,Agarwal:2019gjk,Marolf:2020xie,Mousatov:2020ics,Iliesiu:2020qvm,Kim:2020cds,Verlinde:2020upt,Chen:2020wiq,Gautason:2020tmk,Anegawa:2020ezn,Giddings:2020yes,Hashimoto:2020cas,Sully:2020pza,Hartman:2020swn,Agon:2020fqs,Hollowood:2020cou,Alishahiha:2020qza,Banks:2020zrt,Geng:2020qvw,Zhao:2019nxk,Chen:2019uhq,Bousso:2019ykv,Akers:2019nfi,Balasubramanian:2020hfs,Almheiri:2019psy,Laddha:2020kvp,Saraswat:2020zzf,Krishnan:2020oun,Chen:2020uac}.

On the other hand, a Page curve was also obtained in a simple wormhole model~\cite{Akers:2019nfi} as well as a End-of-the-World (EOW) brane model~\cite{Balasubramanian:2020hfs} in three dimensions, where no quantum fields are involved. This indicates that a Page curve could be seen in holographic models of black hole evaporation even in the classical level. In particular, the applications of Ryu-Takayanagi (RT) or Hubeny-Rangamani-Takayanagi (HRT) formula to (multi-boundary) wormholes are enough to give a Page curve in these models.

While most of the above studies concern the entanglement entropy, in this paper we want to measure the correlation in mixed states during black hole evaporation, for which the entanglement entropy does not work since it is known that entanglement entropy is a faithful measure only for bipartite pure states~\cite{Takayanagi:2017knl,Umemoto:2018jpc}. Therefore, we study another measure called reflected entropy $S_R$, which has been proposed recently~\cite{Dutta:2019gen} (also see~\cite{Bao:2019zqc,Chu:2019etd,Akers:2019gcv,Kudler-Flam:2020url,Moosa:2020vcs,Bueno:2020vnx} for further development) based on canonical purification of a given density matrix $\rho_{AB}$,

\begin{equation}
S_R(A:B) := S(AA^*)_{\sqrt{\rho_{AB}}}\ ,
\end{equation} where $|\sqrt{\rho_{AB}}\rangle$ is a canonically purified state in the doubled Hilbert space and $S(AA^*)$ is von Neumann entropy.

In particular, let us consider a general model of black hole evaporation, which could include several-side black holes (such as eternal black hole) and more than one reservoirs of radiation. Assuming the Hilbert spaces of the whole system is factorized, we want to measure the reflected entropy between any two of the subsystems. This basically includes the reflected entropy between black hole and radiation, the reflected entropy between black hole and black hole and the one between radiation and radiation.

Below we summarize the main results of this paper.

We start from the analysis of a simple 3-side wormhole model, where we find that the holographic dual of reflected entropies are actually island cross sections. We plot the curve of reflected entropy as a function of time and find that: the reflected entropy between black hole and radiation basically follows the behavior of Page curve but with a later transition time (compared with Page time), the reflected entropy between radiation and radiation increases at early time and then saturates. We then move to the EOW brane model, similar behaviors of reflected entropy are found.

We revisit the holographic reflected entropy in AdS/CFT, aiming to find a similar formula as quantum extremal surface (QES) formula, for the reflected entropy. We conjecture a quantum extremal cross section (QECS) formula for exact reflected entropy and argue that it is a direct consequence of QES formula due to a $Z_2$ symmetry of canonical purification.

Similar to the QES formula of Von Neumann entropy, the QECS formula of reflected entropy is expected to work for more general gravitational systems. Having this in mind, we move to the eternal black hole+CFT model. There we find a generalized formula for reflected entropy between left radiation and left black hole, provided that the right side can be treated as the canonical purification of the left. This formula can be further generalized to the black hole-black hole reflected entropy and the radiation-radiation reflected entropy. We plot the numerical results by employing these formulas and particularly find that the reflected entropy between the left black hole and the right black hole decreases to zero during the evaporation process. The behaviors of entropy curves as functions of time agree with the earlier wormhole models.

This paper is organized as follows. In Section 2, we analyze 3-side wormhole model and compute reflected entropy using the Ryu-Takayanagi formula for multi-boundary states and gluing procedure. In particular we establish the computation of cross sections in covering space. In Section 3, we compute reflected entropies in the 3d EOW brane model by extending the covering space method in previous section. We revisit holographic reflected entropy with quantum corrections in Section 4 and propose a quantum extremal surface for reflected entropy. In Section 5 we analyze 2d eternal black hole plus CFT model and find generalized formula for reflected entropies between black hole and radiation, black hole and black hole, and radiation and radiation. We conclude and discuss future questions in Section 6.

While this paper is in preparation we got to know an independent work by Venkatesa Chandrasekaran, Masamichi Miyaji and Pratik Rath~\cite{Miyaji}, which will appear together with this paper.

\section{Reflected entropy in simple wormhole model\label{sec2}}

\begin{figure}[htbp]
		\centering 
		\subfigure[$m_1+m_2<m_3$]{ 
			\begin{minipage}{6cm}
			\centering 
			\includegraphics[scale=0.4]{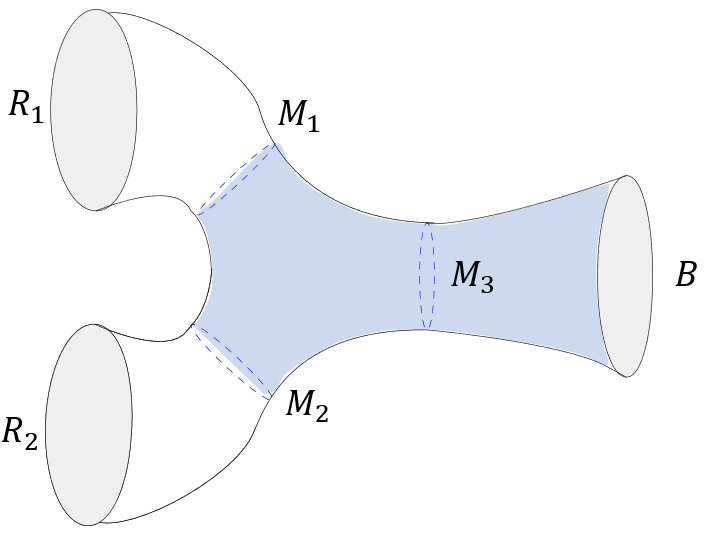} 
			\end{minipage}
		}
		\subfigure[$m_1+m_2>m_3$]{ 
			\begin{minipage}{6cm}
			\centering 
			\includegraphics[scale=0.4]{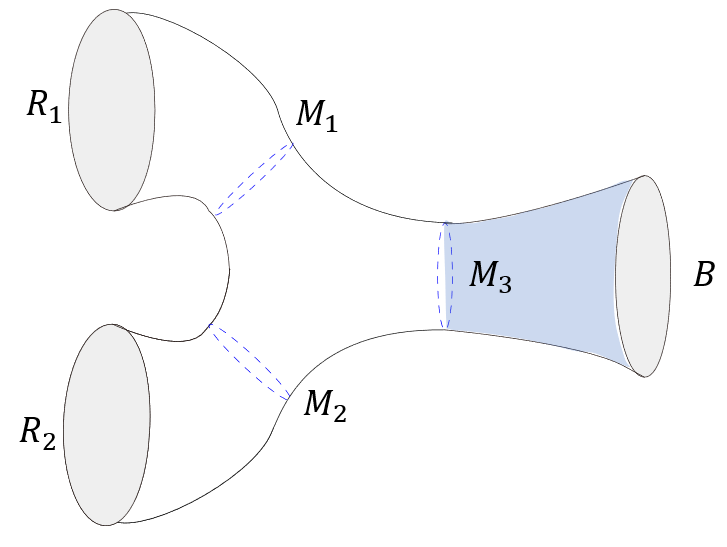} 
			\end{minipage}
		}
		\caption{The 3-side wormhole has three asymptotic boundaries $R_1$, $R_2$ and $B$. The entanglement wedge of $B$ is the blue shaded region in the figure. We have $m_1+m_2<m_3$ at early time of the evaporation and $m_1+m_2>m_3$ at late time.} 
		\label{3side} 
	\end{figure}

We follow~\cite{Akers:2019nfi} to model the black hole evaporation using a multi-boundary wormhole (Fig.\ref{3side}). We can imagine splitting the Hawking radiation into $n$ different parts. The state of the total system containing the black hole and the radiation can be written as \\
\begin{equation}\label{entanglement}
\ket{\Psi}= \sum_{i_1\cdots i_n} c_{i_1\cdots i_n}\ket{i_1}_{R_1}\otimes \cdots \ket{i_n}_{R_n}\cdots \ket{\Psi_{i_1\cdots i_n}}_B
\end{equation} \\
where \begin{math}\ket{i}\end{math} is the Hawking radiation and \begin{math}\ket{\Psi_{i_1\cdots i_n}}\end{math} is the black hole state. For convenience, let us consider the simplest case in which the radiation is split into two parts. Holographically, we can view the CFT states on the boundary $R_1$ and $R_2$ (Fig.\ref{3side}) as two parts of the Hawking radiation, which are entangled with the CFT states of the black hole on the boundary $B$. The two parts of emitted radiation and the original black hole are connected through a three-side wormhole. While~\cite{Akers:2019nfi} increases the number of legs of the radiation to simulate the black hole evaporation, here we choose to keep the number of the legs fixed but increase the size of the horizons corresponding to the radiation states. 

\par
For simplicity, we take the two horizons $M_1$ and $M_2$ to have the same size $m_1=m_2$. $M_1$ and $M_2$ are actually RT surfaces of $R_1$ and $R_2$ respectively. Let the length of the horizon $M_3$ of the original black hole be $L_0$ and $m_1=m_2=0$ as the initial condition. Since the ADM energy is conserved during the evaporation process, as $m_1$ and $m_2$ increases, $m_3$ will decrease. The energy-entropy relation of black hole excitation in AdS$_3$ is
\begin{equation}\label{ee_relation}
S = 2\pi\sqrt{\frac{cE}{3}},
\end{equation} and therefore at any moment during evaporation, the horizon length of the black hole $B$ is determined by
\begin{equation}\label{energy-conservation}
m_3 = \sqrt{L_0^2-2m_1^2}.
\end{equation}
As shown in Fig.\ref{3side}, the entanglement wedge of radiation covers the shared interior after the transition between different RT surfaces, therefore the shared interior is considered as the island in this model.
	
	\begin{figure}[htbp]
	\centering
	\includegraphics[scale=0.4]{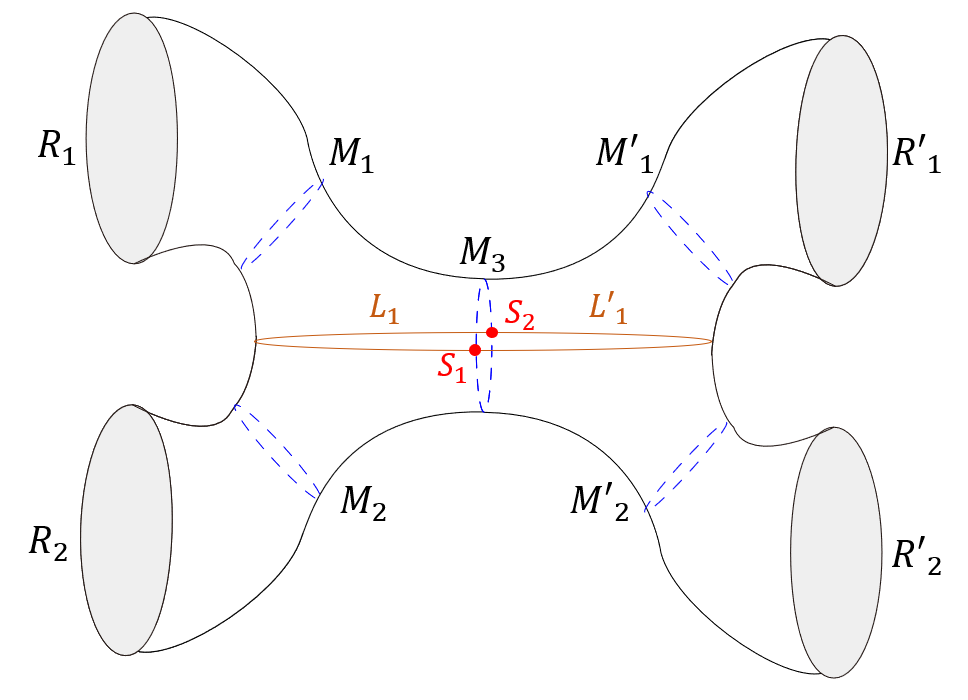}\\
	\caption{The purified geometry is a 4-side wormhole which is symmetric with respect to the horizon $M_3$. The brown geodesic $L_1 \cup L_1'$ intersects $M_3$ at two points $s_1$ and $s_2$. The minimal cross section can be obtained by minimizing the length of $L_1 \cup L_1'$ or equivalently, $L_1$, over $s_1$ and $s_2$.}
	\label{4side}
	\end{figure}
\subsection{Reflected entropy between $R_1$ and $R_2$} \label{sec3sideRR}
In order to compute the reflected entropy between $R_1$ and $R_2$, we need to trace out $B$. Holographically, this means we have to remove the entanglement wedge of $B$, take two copies of the remaining geometry and glue them together. The RT surface of $B$ is the smaller one of $M_1\cup M_2$ and $M_3$ (Fig.\ref{3side}). At early time when the radiation horizon is small, the RT surface of $B$ is $M_1\cup M_2$. As time goes by, $M_3$ will dominate. Therefore, when $m_1+m_2<m_3$, the entanglement wedge of $B$ is the blue shaded region in Fig.\ref{3side}(a). In this case $R_1$ and $R_2$ are disconnected after removing the entanglement wedge. Thus, the reflected entropy between $R_1$ and $R_2$ is simply zero. When $m_1+m_2>m_3$, the entanglement wedge of $B$ is the shaded region in Fig.\ref{3side}(b). After removing it, we glue two copies of the remaining geometry through $M_3$ (Fig.\ref{4side}). The reflected entropy between $R_1$ and $R_2$ now corresponds to the minimal cross section which separates $R_1\cup R_1'$ and $R_2\cup R_2'$. As we can see from Fig.\ref{4side}, we have two options, namely the brown curve $L_1 \cup L_1'$ or the union $M_1\cup M_1'$. To summarize, when $m_1<{m_3\over 2}$, the reflected entropy between $R_1$ and $R_2$ is $0$. When $m_1>{m_3\over 2}$, the reflected entropy is given by
	\begin{equation}\label{min}
	S_R(R_1:R_2) = \text{min} \biggr\{\frac{2m_1}{4G_N}, \frac{2L_1}{4G_N}\biggr\}.
	\end{equation}

	\begin{figure}[htbp]
	\centering
	\includegraphics[scale=0.4]{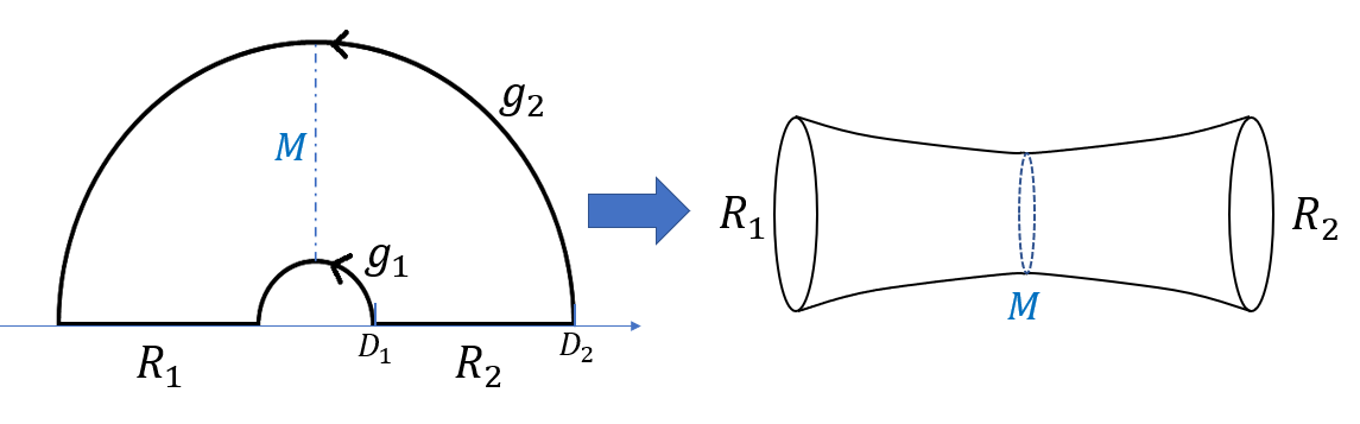}\\
	\caption{The covering space construction of a 2-side wormhole. The geodesics $g_1$ and $g_2$ are identified. The blue dashed circle $M$ is the causal horizon and $R_1$ and $R_2$ are two asymptotic boundaries.}
	\label{2side}
	\end{figure}
\par
Our next step is to compute geodesics in (\ref{min}) in a dynamical evaporation process. First we will give a brief review of the construction of multi-boundary wormhole. For a more detailed discussion on this topic, we refer to~\cite{Caceres:2019giy,Balasubramanian:2014hda}. Multi-boundary wormholes can be viewed as a geometry where boundary CFTs are connected by a wormhole. All the boundaries have independent Hilbert spaces. We can construct the multi-boundary wormhole geometries by quotienting the hyperbolic upper half space $\mathbb{H}^2$ by an isometry subgroup $\Gamma \subset PSL(2,\mathbb{R})$. The action of $\Gamma$ identifies a pair of geodesics on $\mathbb{H}^2$ and the quotient space $\mathbb{H}^2/\Gamma$ is just the multi-boundary wormhole geometry, which can be interpreted as a time-reflection-symmetric slice of a 2+1d geometry with the metric
	\begin{equation}\label{metrics}
	\text{d}s^2 = -\text{d}t^2 + l^2 \cos^2 \frac{t}{l}\text{d}\Sigma,
	\end{equation}
where d$\Sigma$ is the metric inherited from $\mathbb{H}^2$. As an example, we can create a two-side wormhole by identifying two geodesics $g_1$ and $g_2$ in Fig.\ref{2side}
	\begin{equation}\label{geodesics}
	g_1(\lambda) = D_1 e^{i\lambda}, \quad
	g_2(\lambda) = D_2 e^{i\lambda},
	\end{equation}
where $\lambda \in [0, \pi]$ is the curve parameter. The isometry group is made up of one single element $\gamma_1$, which can be written in $SL(2,\mathbb{R})$ form as

	\begin{equation}\label{gamma1}
	\gamma_1 = 
	\left(
	\begin{array}{ll}
	\sqrt{\frac{D2}{D1}} & 0\\
	0 & \sqrt{\frac{D1}{D2}} \\
	\end{array}
	\right).
	\end{equation}
This transformation can also be written in a simple form as $\gamma_1(z)=\frac{D_2}{D_1}z$, where $z$ is complex coordinate. It sends points on the smaller semicircle $g_1$ to the larger semicircle $g_2$. Note that the blue dashed vertical line (geodesic) is invariant under this transformation. It is taken as the defining feature of the horizons in multi-boundary wormhole, i.e. they are geodesics invariant under a combination of the generators in the isometry group $SL(2,\mathbb{R})$. The geodesics in covering space are either vertical lines or semicircles with centers on the horizontal axis. The region surrounded by $g_1$ and $g_2$ is one fundamental region of the quotient space. Intuitively, it can be viewed as an unfolded diagram of the 2-side wormhole in the right of Fig.\ref{2side}. 

	\begin{figure}[htbp]
	\centering
	\includegraphics[scale=0.4]{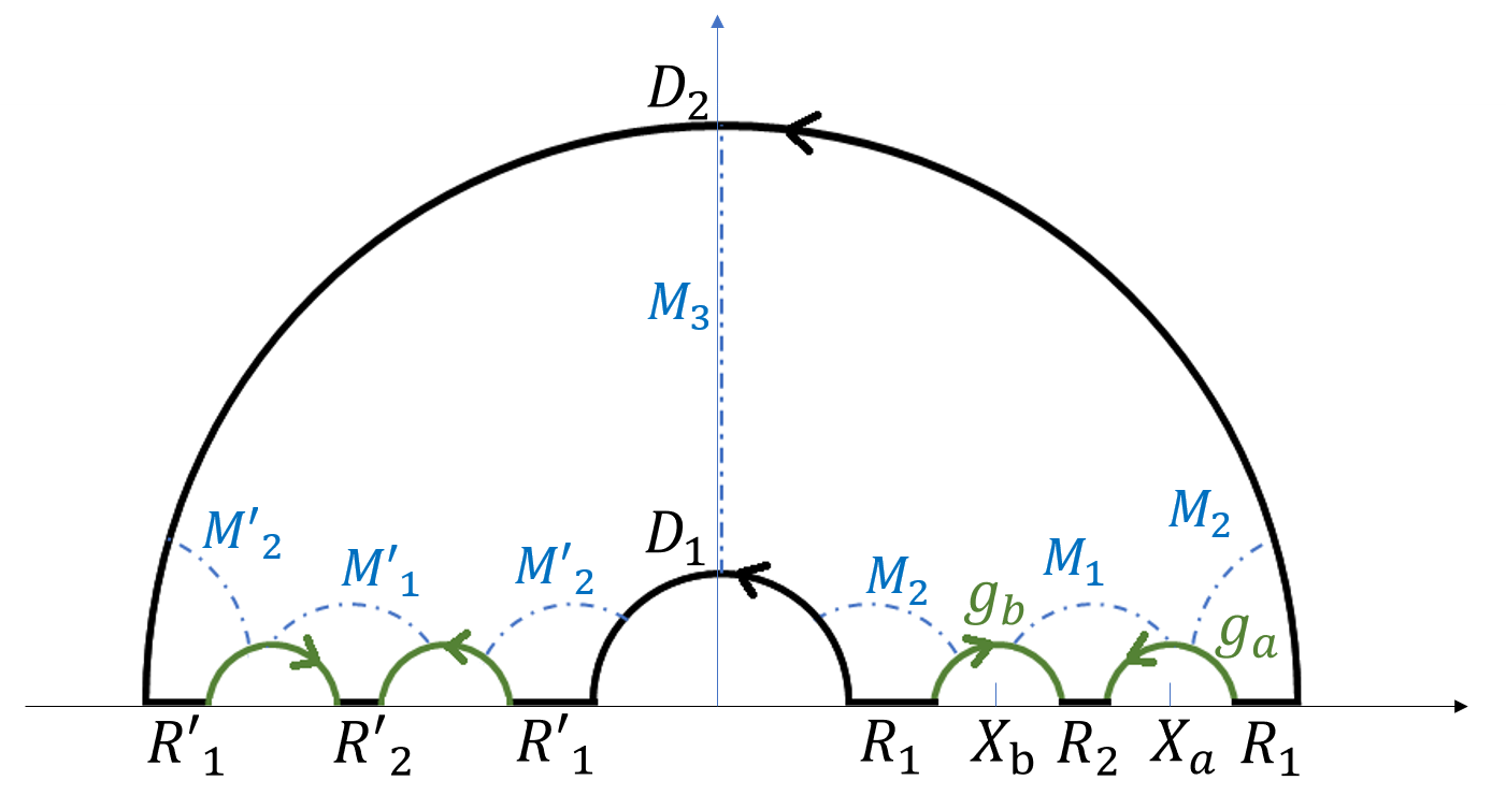}\\
	\caption{A fundamental region of the quotient space of the 4-side wormhole in Fig.\ref{4side}. It is symmetric with respect to the horizon $M_3$. The geodesics $g_a$ and $g_b$ are identified in an orientation-reversed way.}
	\label{covering}
	\end{figure}

\par
One can introduce more legs in the wormhole by removing more half-disks in the covering space and identifying the semicircles in an orientation reversing way. For instance, we remove four semicircles in the covering space (Fig.\ref{covering}) to construct the 4-side wormhole in Fig.\ref{4side}. The causal horizons corresponding to different boundaries are blue dashed curves. We identify the semicircle $g_b$ with another semicircle $g_a$ in a reverse orientation, which can be written as

	\begin{equation}\label{geodesics}
	g_a(\lambda) = X_a + D_a e^{i\lambda},
	\end{equation}
	\begin{equation}\label{geodesics}
	g_b(\lambda) = X_b - D_b e^{-i\lambda}.
	\end{equation}
The generator which identifies $g_a(\lambda)\sim g_b(\lambda)$ is
	
	\begin{equation}\label{gamma2}
	\gamma_2 = 
	\left(
	\begin{array}{ll}
		\sqrt{D_a} & \frac{X_a}{\sqrt{D_a}}\\
		0	&	\frac{1}{\sqrt{D_a}}\\
	\end{array}
	\right)
	\left(
	\begin{array}{ll}
		0	&	-1\\
		1	&	0
	\end{array}
	\right)
	\left(
	\begin{array}{ll}
		\frac{1}{\sqrt{D_b}} & -\frac{X_b}{\sqrt{D_b}}\\
		0	&	\sqrt{D_b}
	\end{array}
	\right)\ .
	\end{equation}.

The identification between $g_a$ and $g_b$ in the covering space introduces two boundaries $R_1$ and $R_2$ in Fig.\ref{4side}. Due to the apparent $Z_2$ symmetry, we do the same thing in the left part of the covering space (Fig.\ref{covering}), which introduces the other two boundaries $R_1'$ and $R_2'$. Note that the horizons homologous to the respective boundaries of the wormhole are the geodesics which are invariant under the combinations of the isometry group generators. For example, $M_1$, $M_2$ and $M_3$ are geodesics invariant under $\gamma_2$, $\gamma_1\circ\gamma_2^{-1}$ and $\gamma_1$, respectively. The lengths of these horizons are three independent parameters determining the wormhole's size, and they can be expressed in terms of six circle parameters $D_1, D_2, D_a, D_b, X_a, X_b$ in the covering space. For instance, using the metric of the covering space, we can express $m_3$ in terms of circle parameters

	\begin{equation}\label{m3}
	m_3 = \log{\frac{D_2}{D_1}}.
	\end{equation}

\par
\begin{figure}[htbp]
	\centering
	\includegraphics[scale=0.4]{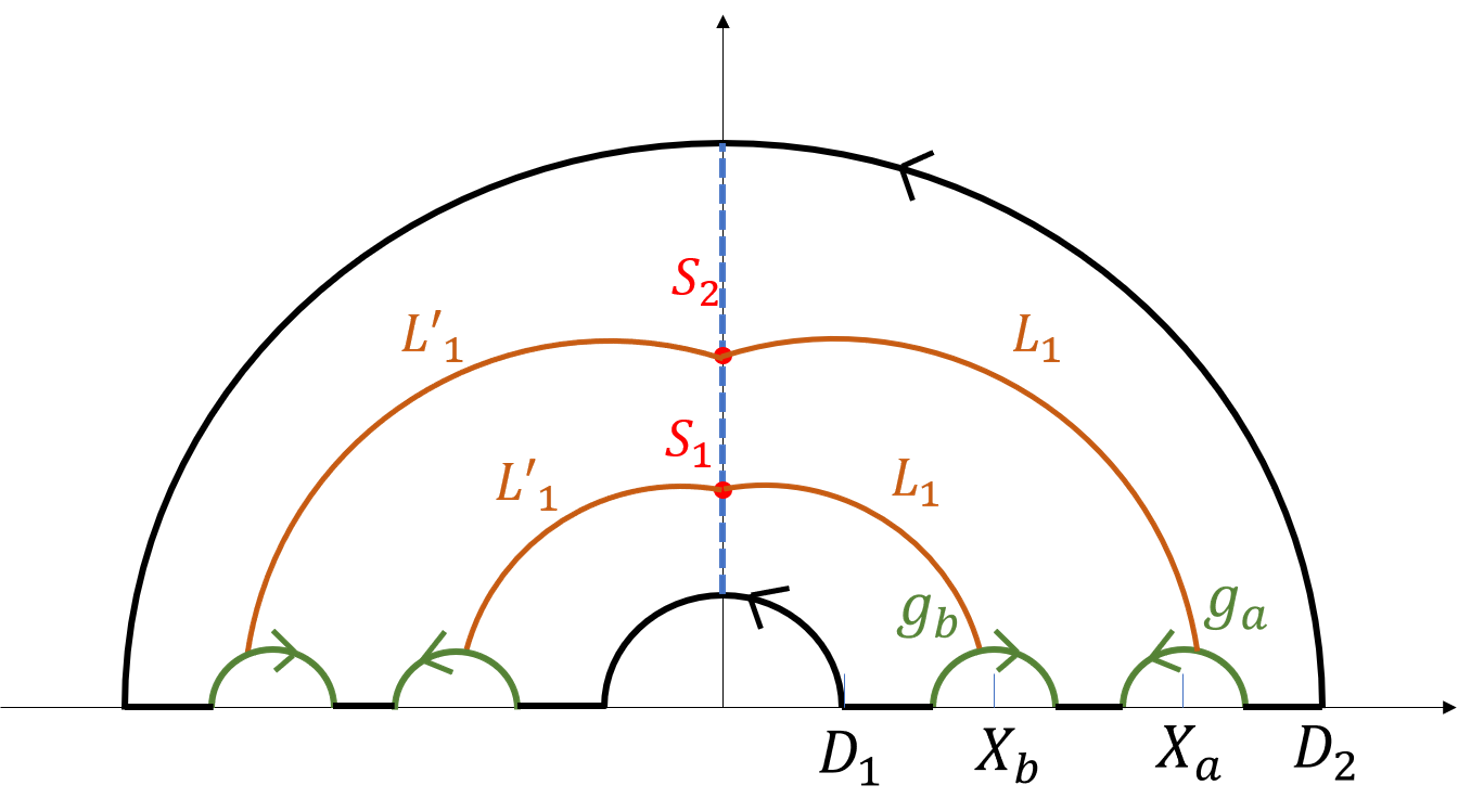}\\
	\caption{The brown geodesic in Fig.\ref{4side} with two intersection points $s_1, s_2$ on the horizon $M_3$ is depicted in this covering space. This covering space has a $Z_2$ symmetry with respect to the vertical dashed blue line $M_3$, so we only have to consider the right half. $L_1$ consists of two arcs. The larger one is part of a large semicircle which is the image of $\gamma_2$ (\ref{gamma2}) applied to the smaller semicircle that includes the smaller arc of $L_1$. The two brown arcs are joint smoothly under the identification of the green semicircles. This type of geodesic is unique once we fix the intersection points $s_1$ and $s_2$, so we can move $s_1,s_2$ to obtain the minimal one.}
	\label{covering2}
	\end{figure}

The lengths of the other two horizons can be calculated in terms of circle data too. The explicit formulas of $m_1$ and $m_2$ can be found in~\cite{Balasubramanian:2020hfs}. One can also use the covering space to compute the cross section $L_1 \cup L_1'$ in Fig.\ref{4side}. It is simply the brown curve in Fig.\ref{covering2} which has two intersection points $s_1, s_2$ on the horizon $M_3$. Since the two sides of the geometry are $Z_2$ symmetric with respect to $M_3$, the total length of the cross section is twice the length of $L_1$. The explicit formula of $L_1(s_1,s_2)$ is given in Appendix \ref{L1-L2-formulas}. Note that we need to minimize $L_1(s_1,s_2)$ over $s_1,s_2$ in order to obtain the minimal cross section. Conclusively, geodesics $m_1$ and $L_1$ in equation (\ref{min}) can both be calculated in terms of the circle data in the covering space. 

\par
Since we impose the energy conservation constraint (\ref{energy-conservation}) in the dynamical evaporation process and we have taken $m_1=m_2$, there is only one free parameter left in the 3-side wormhole model. In the covering space however, it is worth noting that there is always a redundancy because there are totally six circle parameters. Therefore, we fix the circle data in covering space by expressing them in terms of only one free parameter. Since we set $m_1 = m_2$, we can equivalently equate the eigenvalues of $\gamma_2$ and $\gamma_1\circ\gamma_2^{-1}$ and we get the following relation
	\begin{equation}\label{xaxb}
	X_a = \mu X_b,\\
	\end{equation}
where $\mu:=\sqrt{\frac{D_2}{D_1}}$. We then fix the remaining circle parameters by the following setting$\footnote{As explained in~\cite{Balasubramanian:2020hfs}, there are different choices to fix the circle data. The point is that one should make such a choice that keeps $D_1<X_b-D_b<X_b+D_b<X_a-D_a<X_a+D_a<D_2$ for any $\mu>1$, which ensures all the curves are in the fundamental region (Fig.\ref{covering2}).}$

	\begin{equation}
	D_2 = \mu^2 D_1,
	\end{equation}
	\begin{equation}
	X_a = \frac{\mu^2+\mu}{2} D_1,
	\end{equation}
	\begin{equation}
	X_b = \frac{\mu+1}{2} D_1,
	\end{equation}
	\begin{equation}
	D_a = \mu D_b.
	\end{equation}
Here $D_1$ can be an arbitrarily positive constant. Now for a given $m_3$ at any moment, we can solve all the circle parameters using the relations above together with (\ref{energy-conservation}), (\ref{m3}) and the formula of $m_1$. Then we can use the formula of $L_1(s_1, s_2)$ (Appendix \ref{L1-L2-formulas}) to work out the half cross section $L_1$ and minimize it over $s_1,s_2$. 
	\begin{figure}[htbp]
	\centering
	\includegraphics[scale=0.4]{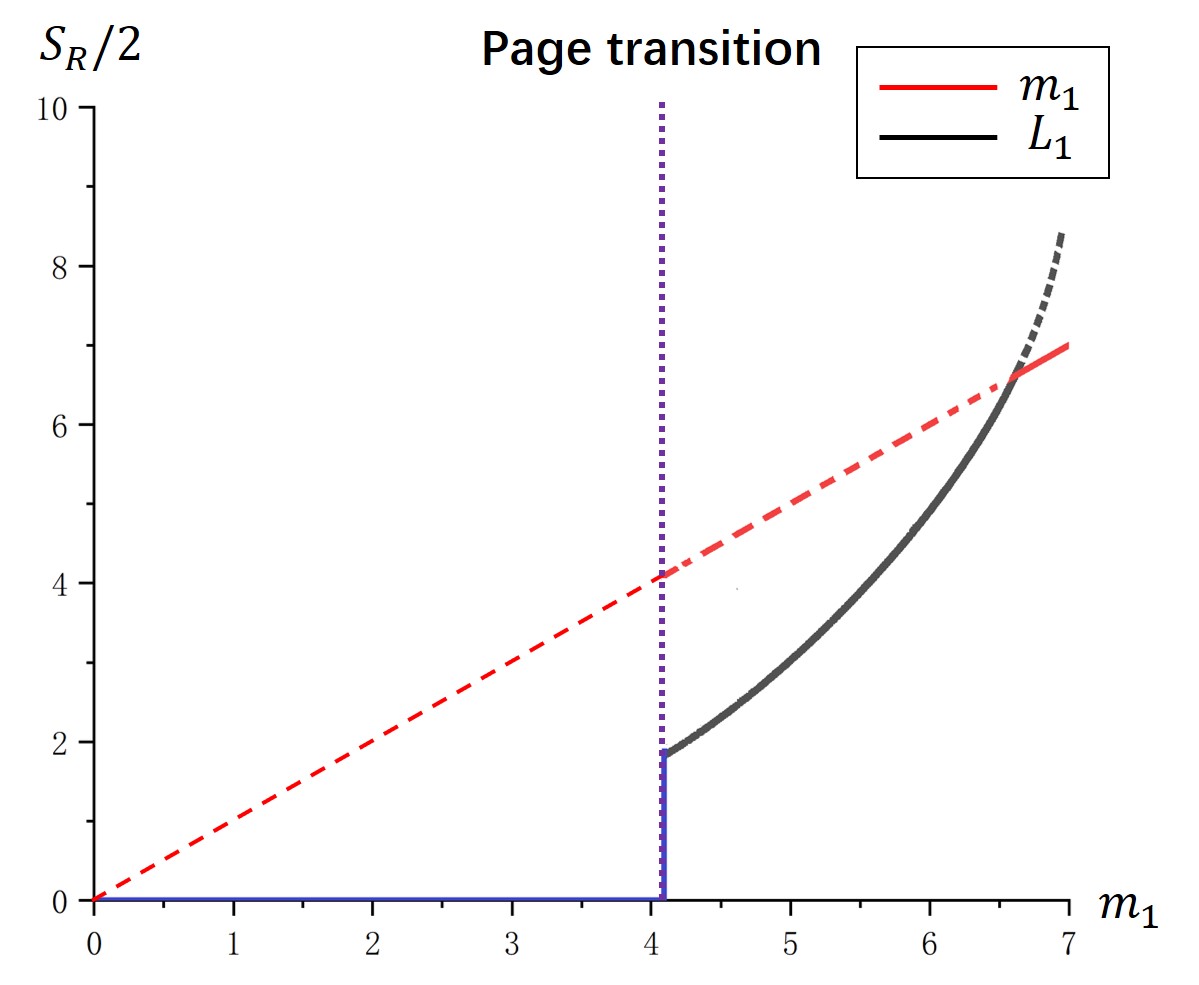}\\
	\caption{We set $4G_N=1$ and plot the half of the reflected entropy as a function of the horizon length $m_1$. The solid curves are half the reflected entropy between $R_1$ and $R_2$. The purple vertical dashed line characterizes the Page transition when $2m_1=m_3$.}
	\label{3sideRR}
	\end{figure}

\par
When we set $L_0=10$ and $4G_N=1$, the reflected entropy between $R_1$ and $R_2$ as a function of $m_1$ during the whole evaporation process is plotted in Fig.\ref{3sideRR}. The red curve is the length of $M_1$ and the black curve is the length of $L_1$. The reflected entropy picks up the smaller one of the two competing curves, i.e. the lower solid curve in Fig.\ref{3sideRR}. We can see that the reflected entropy between radiation and radiation jumps from zero to a positive value at $m_1\approx4$ when the island appears (i.e. $2m_1>m_3$), and then it keeps increasing as the growth of the amount of radiation. It goes through a phase transition from the black curve $L_1$ to the red curve $m_1$ near $m_1=6.65$. At the end of the evaporation, with $m_3\approx0$, the 3-side wormhole becomes a cylinder which has only two asymptotic boundaries $R_1$ and $R_2$, so the reflected entropy between them saturates at the final value, $m_1 \approx 7$. 

	\begin{figure}[htbp]
	\centering
	\includegraphics[scale=0.4]{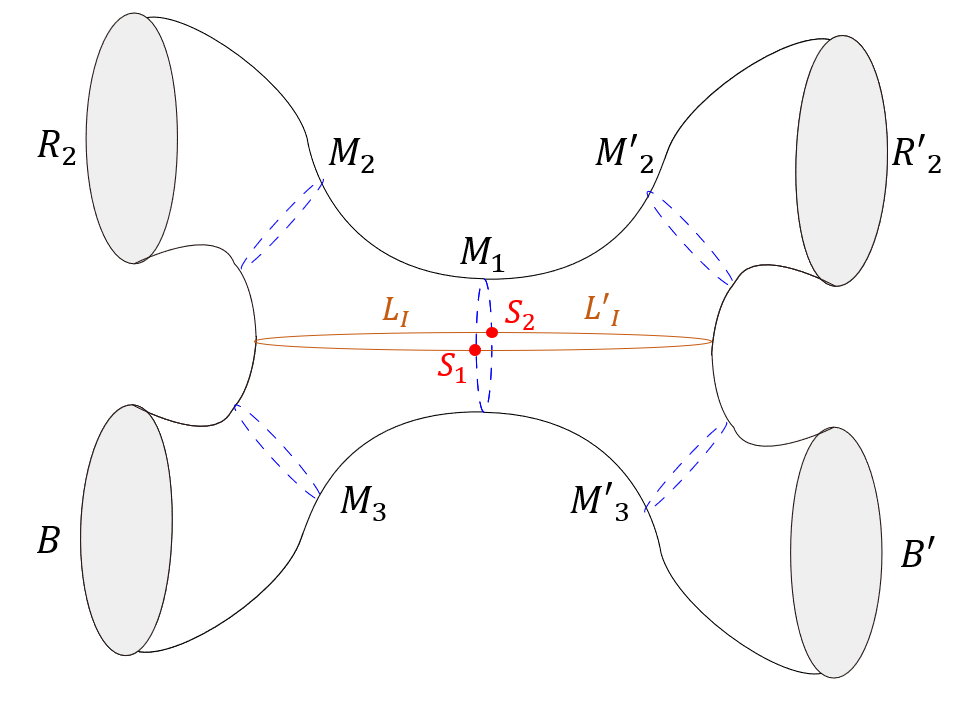}\\
	\caption{The purification is similar to Fig.\ref{4side}. The only different point is that here we remove the entanglement wedge of $R_1$ and purify $R_2$ and $B$. Note that the reflected entropy between $R_2$ and $B$ corresponds to twice the length of the minimal geodesic of $M_2$, $L_I$ and $M_3$.}
	\label{4side2}
	\end{figure}
\subsection{Reflected entropy between $R_2$ and $B$}
\par
Now we compute the reflected entropy between the radiation $R_2$ and the black hole $B$. We remove the entanglement wedge of $R_1$ and glue the two copies of remaining geometry through $M_1$ (Fig.\ref{4side2}). We have three options of geodesics which separates $R_2\cup R_2'$ and $B\cup B'$ in Fig.\ref{4side2}, namely $M_2\cup M_2'$, $M_3\cup M_3'$ or $L_I \cup L_I'$. The reflected entropy between $R_2$ and $B$ is given by
\begin{equation}
S_R(R_2:B) = {1\over 4G_N}\,\text{min}\biggr\{2m_2, 2m_3, 2L_I\biggr\}\ .
\end{equation}
$L_I$ can be computed by the brown curve in Fig.\ref{covering3}, which has two endpoints on $M_1$. We give the explicit formula of $L_I$ in Appendix \ref{LI-section}. Under the same parameter setting in computing the reflected entropy between $R_1$ and $R_2$, we plot the reflected entropy between $R_2$ and $B$ as a function of $m_1$ in Fig.\ref{3sideRB}. 

	\begin{figure}[htbp]
	\centering
	\includegraphics[scale=0.3]{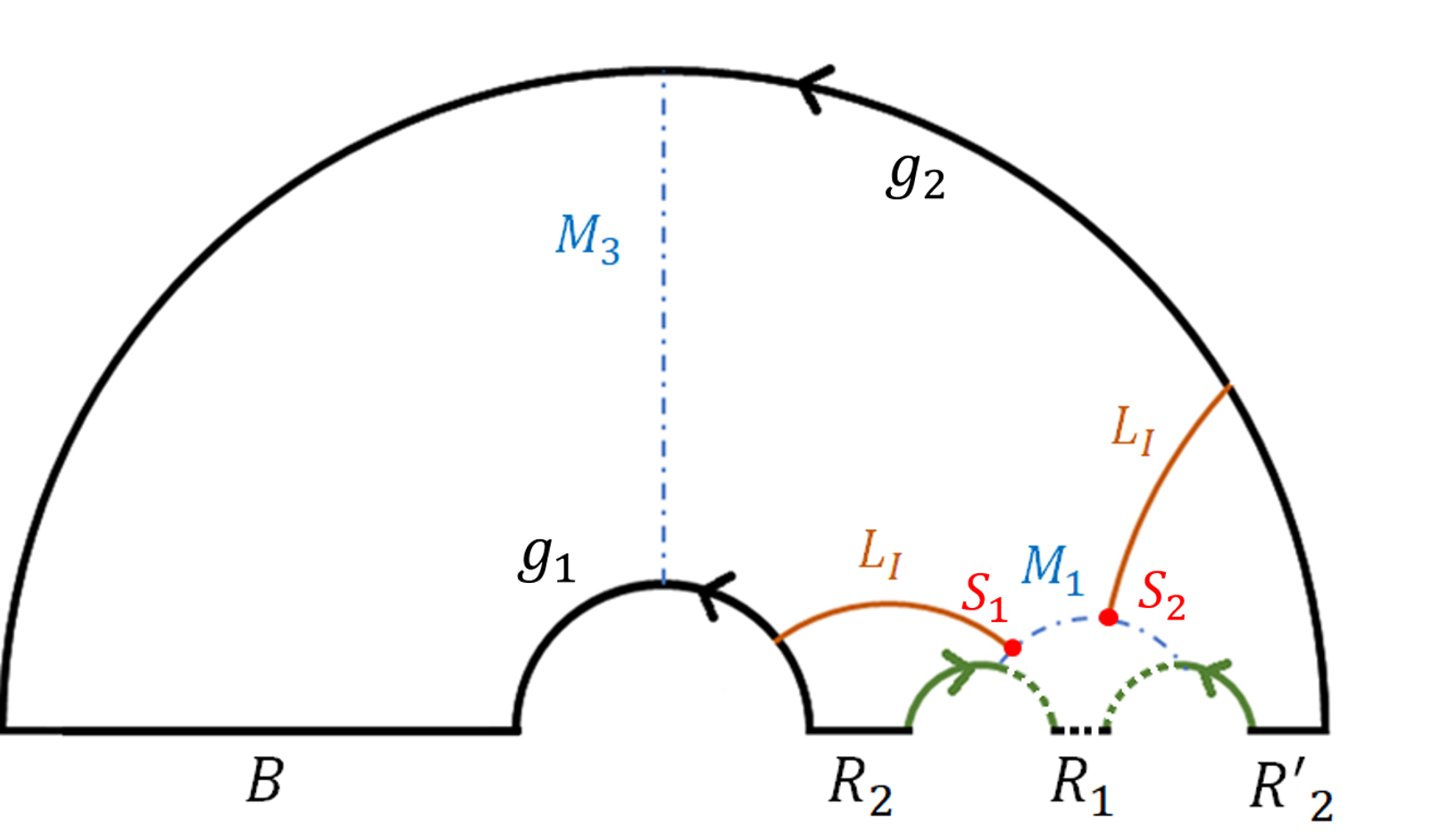}\\
	\caption{This is a covering space depiction of the left half side of Fig.\ref{4side2}. Since we remove the entanglement wedge of $R_1$ in the original 3-side wormhole, the region surrounded by $M_1$ and dashed lines is thrown away in this fundamental region. The depiction of $L_I$ in Fig.\ref{4side2} in this covering space consists of two brown arcs with intersection points on $M_1$. The left arc can be transformed by $\gamma_1$ (\ref {gamma1}) to the outside of the fundamental region and smoothly connected to the right end of the right arc. The locations of the two intersection points on $M_1$ uniquely determine this type of geodesic arcs.}
	\label{covering3}
	\end{figure}

	\begin{figure}[htbp]
	\centering
	\includegraphics[scale=0.4]{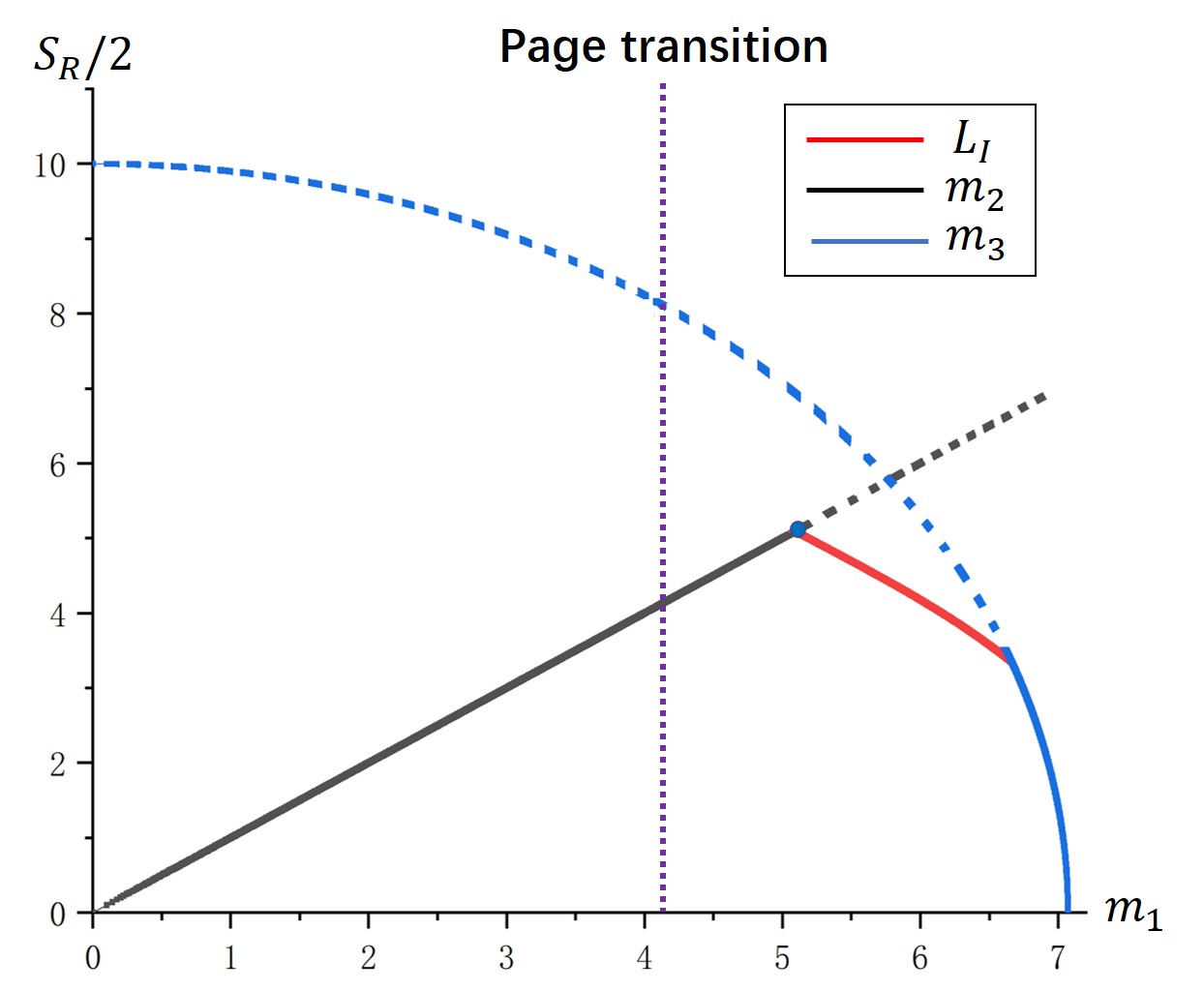}\\
	\caption{The reflected entropy between $R_2$ and $B$ is the minimum of these curves, i.e. the solid curve here. We can see that the transition point of reflected entropy between $R_2$ and $B$ (marked by the blue dot) is later than the Page time.}
	\label{3sideRB}
	\end{figure}

\par
The reflected entropy between $R_2$ and $B$ picks up the lowest curve among the three at any moment of time in the evaporation process. Note that the reflected entropy goes through a phase transition at the intersection point of the black curve and the red curve in Fig.\ref{3sideRB}. It increases first and then goes down to zero at the end of the evaporation. The purple dashed vertical line characterizes the Page transition where $2m_1=m_3$. Note that the transition time for reflected entropy is later than the Page time. 

\section{Reflected entropy in 3d EOW brane model\label{sec3}}
\subsection{Review of the model}
\begin{figure}[htbp]
	\centering
	\includegraphics[scale=0.4]{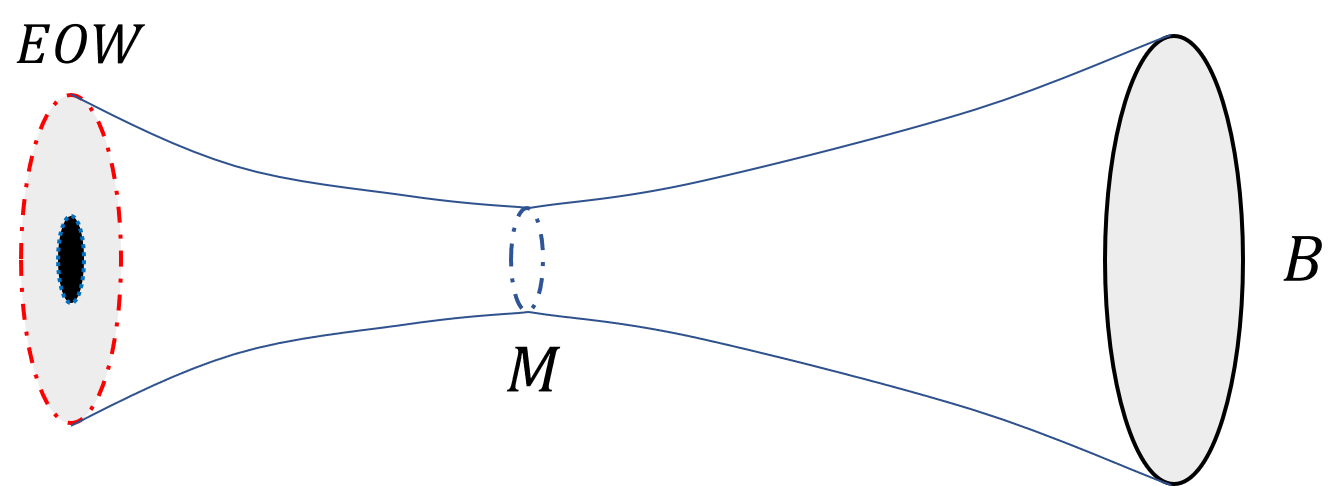}\\
	\caption{The left side of the eternal BTZ black hole is truncated and replaced by an EOW brane (the red dashed circle) with an internal structure which matches that of a 2d CFT representing the interior partner of Hawking radiation. When the CFT is maximally entangled with the radiation, it is in a thermal state and its holographic dual is a black hole (the black region within the disk) . Then we replace the EOW brane with this black hole and glue it to the original geometry at the location of the brane. Note that the entropy of the brane is proportional to the area of the black hole horizon.}
	\label{inception}
	\end{figure}

In~\cite{Balasubramanian:2020hfs}, a different model including an End-of-the-World (EOW) brane for black hole evaporation has been proposed, inspired from two-dimensional JT gravity+EOW model~\cite{Penington:2019kki}. The EOW brane truncates the interior of an eternal black hole in AdS$_3$ and describes the interior partners of the Hawking radiation. The authors in~\cite{Balasubramanian:2020hfs} introduce a brane CFT and then replace it with its holographic dual, which fills in the EOW brane and gives a complete 3D geometry. Consider a situation where the brane CFT states are maximally entangled with the radiation quanta outside the black hole. In this case, the brane CFT will be thermal and its holographic dual will contain a black hole (Fig.\ref{inception}). So there will be two horizons in the whole geometry. One is the original black hole horizon and the other is in the so called \textit{Inception Geometry}, i.e. the holographic dual of the brane CFT.

\begin{figure}[htbp]
	\centering
	\includegraphics[scale=0.4]{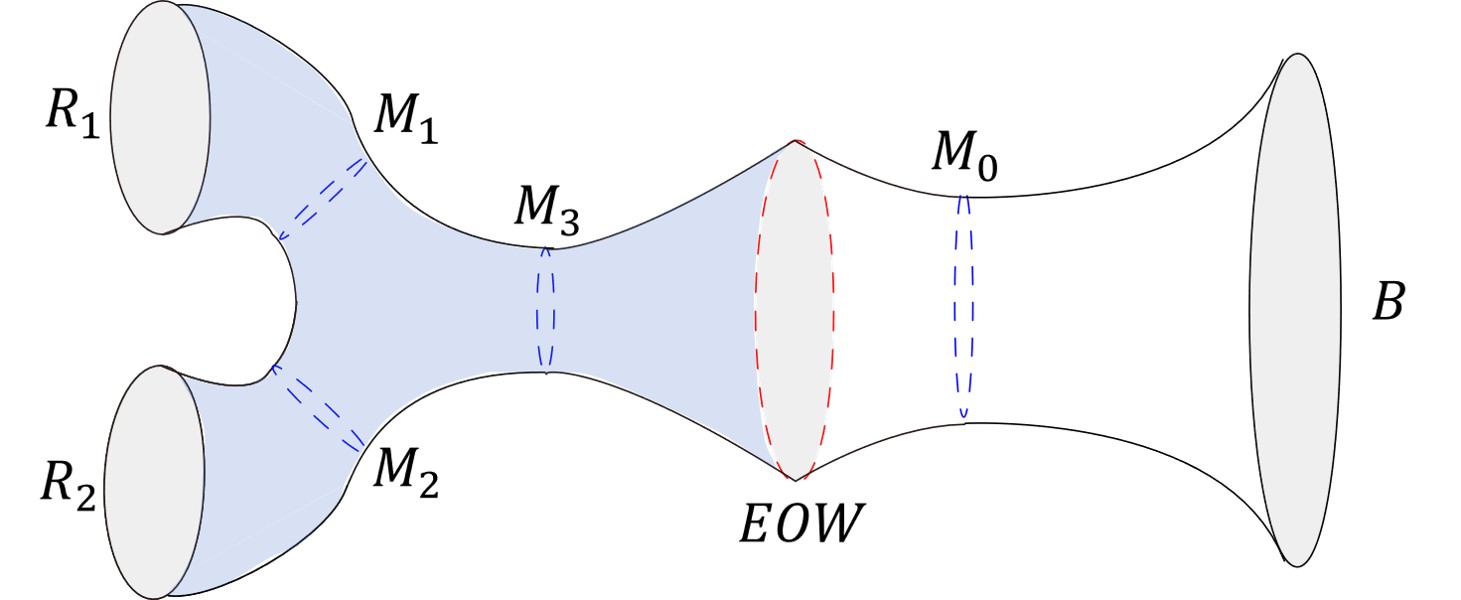}\\
	\caption{We purify the black hole in the inception geometry with a two-boundary wormhole. The two asymptotic boundaries $R_1$ and $R_2$ are two parts of Hawking radiation.}
	\label{eow3side}
	\end{figure}

We follow (\ref{entanglement}) to split the Hawking radiation into $n$ parts and make them entangled with the brane CFT states. To do so, one can purify the black hole in the inception geometry with an auxiliary system which is naturally identified with the Hawking radiation in (\ref{entanglement}). The purified inception geometry can be viewed as a multi-boundary wormhole connecting $n$ asymptotic boundaries. The inception geometry and the real geometry are glued through the EOW brane (Fig.\ref{eow3side}). In order to keep the gluing surface real and non-singular, $m_3\leq m_0$ all the time. The entanglement between the union of radiations and the black hole increases as the black hole evaporates, so one can tune $m_3$ from zero to $m_0$ to model this process. Note that the mass of the original black hole is fixed, which is different from the energy conservation condition in the simple wormhole model in section \ref{sec2}. It means $M_0$ is fixed during evaporation. We will focus on the case of $n=2$ and compute the reflected entropy (Fig.\ref{eow3side}). Following~\cite{Balasubramanian:2020hfs}, we set $m_1=m_2=m_3$, and only one free evaporation parameter is left.

\par
Note that the physical parameters of the original and the brane CFT can be different, so their holographic duals can have different AdS radius $l$ and Newton Constants $G_N$. We denote quantities of the inception geometry by a prime. Following~\cite{Balasubramanian:2020hfs} the central charges of the two CFTs $c$ and $c'$ can be fixed, i.e. $3l/2G_N$ and $3l'/2G_N'$ are fixed during the evaporation process. We take the following gluing condition~\cite{Balasubramanian:2020hfs}
\begin{equation}
r_t=\sqrt{\frac{l^2 G_N^2 r_h'^2-l'^2 G_N'^2 r_h^2}{l^2 G_N^2-l'^2 G_N'^2}},
\end{equation}
where $r_t$ is the position of the EOW brane in the original geometry and $r_h$ and $r_h'$ are the radius of the real and inceptional horizon respectively. We also set $r_t=r_h+\alpha(r_h-r_h')$ with $\alpha>0$ following~\cite{Balasubramanian:2020hfs}. After these settings, now all the quantities scale as functions of the inception horizon radius $r_h'$ or equivalently, $m_3$.~\footnote{This setting of parameters also ensures the non-singular brane trajectories in the evaporation process, for further detail discussions we refer to~\cite{Balasubramanian:2020hfs}.}

	\begin{figure}[htbp]
	\centering
	\includegraphics[scale=0.4]{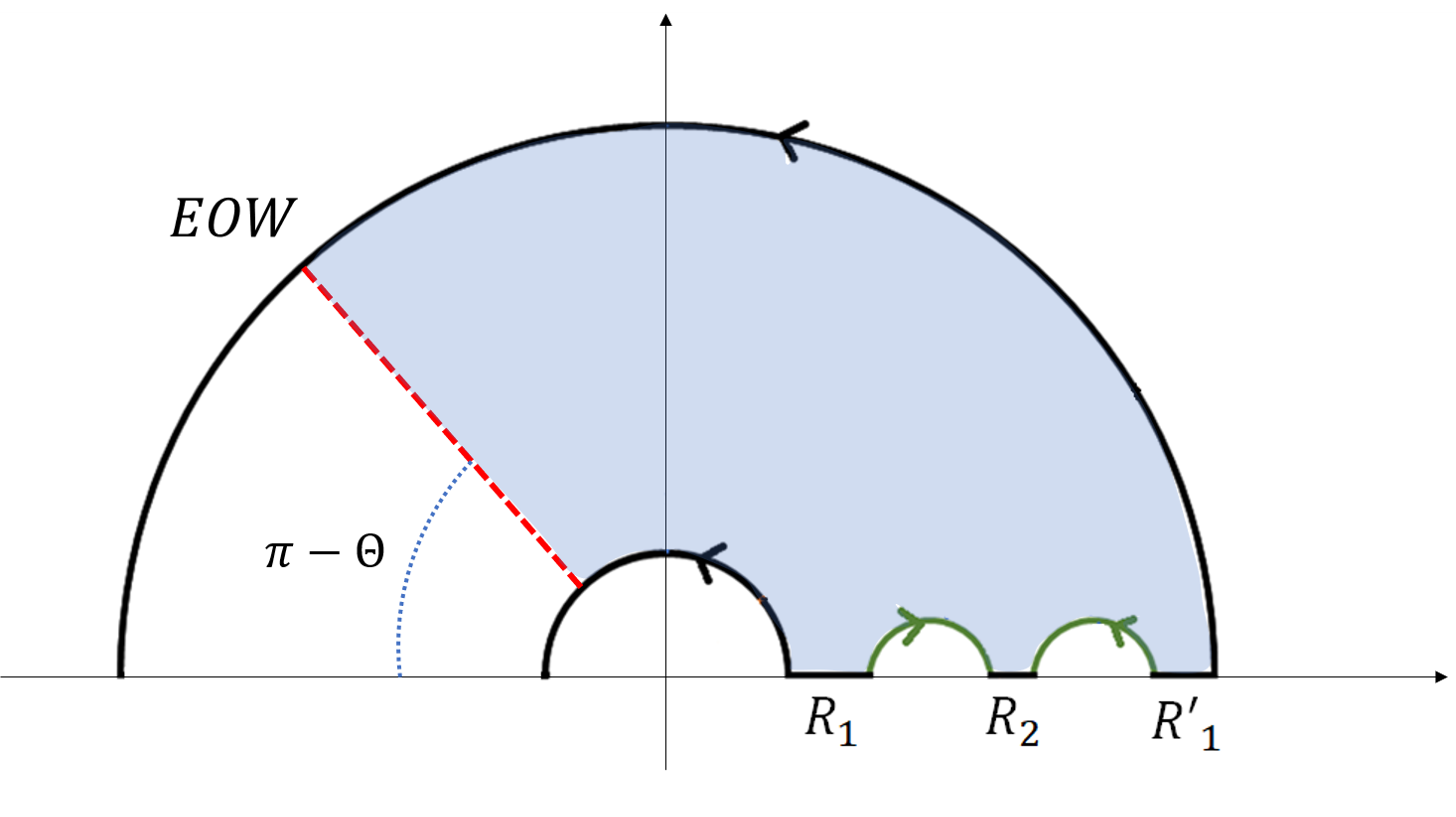}\\
	\caption{A covering space depiction of the inception geometry (the blue shaded region). The red dashed line here is the EOW brane which must match the red dashed circle in Fig.\ref{eow3side}. This covering space can be easily obtained by introducing an EOW brane that truncates the left side of Fig.\ref{covering}. }
	\label{inceptionCovering}
	\end{figure}

\par
One can construct the inception geometry to the left of the EOW brane in Fig.\ref{eow3side} in a covering space (Fig.\ref{inceptionCovering}). This covering space is similar to that of the 3-side wormhole in Fig.\ref{covering}. However, we introduce a red dashed line which truncates the geometry. This line is the EOW brane in Fig.\ref{eow3side}. We now have $m_1=m_2=m_3$, so we can equate the eigenvalues of three group generators $\gamma_1$, $\gamma_2$ and $\gamma_1\circ\gamma_2^{-1}$ and get the following constraints
\begin{equation}
X_a=\mu X_b,\qquad X_b=\frac{\sqrt{D_aD_b}(1+\mu^2)}{\mu(\mu-1)}.
\end{equation}
Then we remove the redundancy of the remaining circle data by expressing them in terms of $\mu$, the only free parameter~\cite{Balasubramanian:2020hfs},
	\begin{equation}
	D_1=\frac{1}{\mu},
	\end{equation}
	\begin{equation}
	D_2=\mu,
	\end{equation}
	\begin{equation}
	D_a=\frac{\mu-1}{\mu},
	\end{equation}
	\begin{equation}
	D_b=\frac{\mu-1}{4\mu}.
	\end{equation}

Again, we emphasize that the fixing of the remaining circle data is not unique, and a good fixing is to maintain $D_1<X_b-D_b<X_b+D_b<X_a-D_a<X_a+D_a<D_2$ for any $\mu>1$, similar to the case in 3-side wormhole. Under this setting of parameters, all the three horizons have the same length $m_1=m_2=m_3=2\log{\mu}$ and the evaporation process can be described by increasing $\mu$ from $\mu=1$ until $m_3=m_0$.
	
\subsection{Reflected entropy between $R_1$ and $R_2$}\label{eowRR-section}
	
	\begin{figure}[htbp]
	\centering
	\includegraphics[scale=0.4]{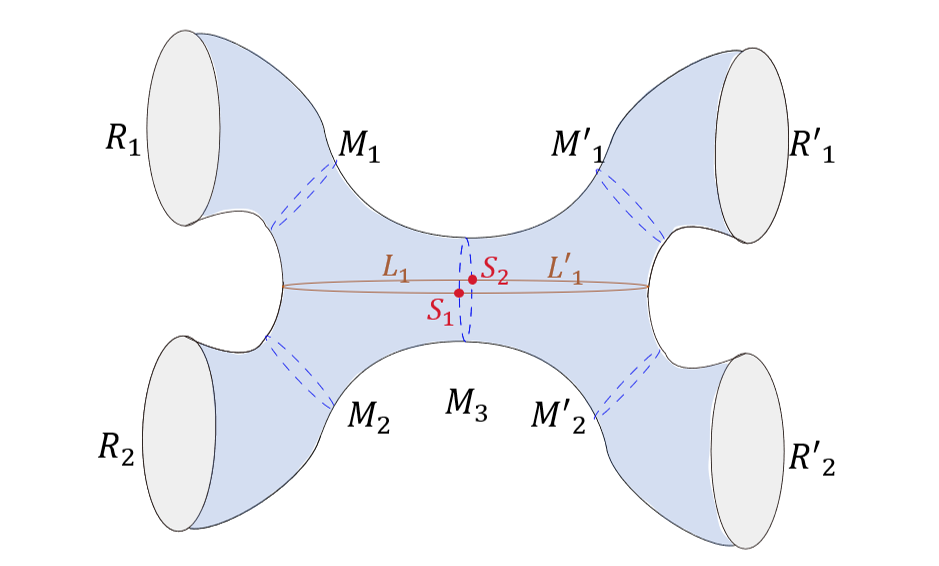}\\
	\caption{The purification of $R_1$ and $R_2$ when removing the right hand side of $M_3$. The cross section $L_1 \cup L_1'$ is the same type as in Fig.\ref{4side}. We use a blue shadow to emphasize that this part is in the inception geometry with a different Newton's constant and a different AdS radius (compared with the original geometry).}
	\label{eowRRphase1Wormhole}
	\end{figure}

\par
We first compute the reflected entropy between the two parts of radiation $R_1$ and $R_2$. Following the method in the previous section, we have to remove the entanglement wedge of the asymptotic boundary $B$, replicate the remaining geometry and glue them together. Note that the RT surface homologous to $B$ is either $M_3$ or $M_0$ (Fig.\ref{eow3side}), so we have two different phases. When $m_3/4G_N'<m_0/4G_N$, $M_3$ is chosen as the RT surface. The purified geometry in this case is shown in Fig.\ref{eowRRphase1Wormhole}. The competing cross sections which split $R_1$, $R_1'$ and $R_2$, $R_2'$ are the curves $L_1 \cup L_1'$, $M_1\cup M_1'$ and $M_2\cup M_2'$. This phase is the same as that in section \ref{sec3sideRR} so we can employ the same formula to compute $L_1$. 

	\begin{figure}[htbp]
	\centering
	\includegraphics[scale=0.4]{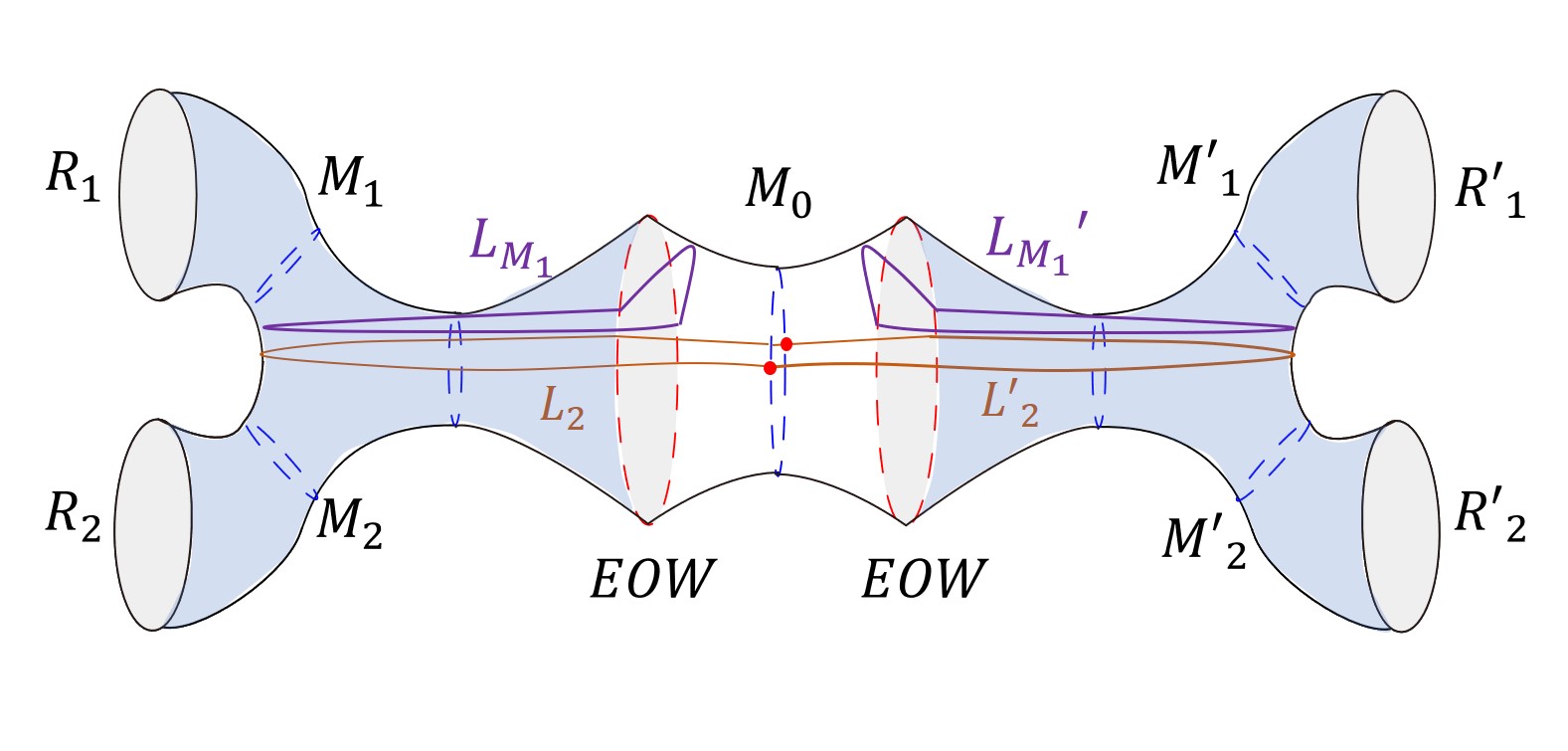}\\
	\caption{The purification geometry glued through $M_0$. The geodesic $L_2$ or $L_2'$ consists of two parts, one in the inception geometry and the other in the original geometry.}
	\label{eowRRphase2Wormhole}
	\end{figure}

\par
When $m_3/4G_N'>m_0/4G_N$, we have to remove the part to the right of $M_0$ in Fig.\ref{eow3side} and glue together two copies of the remaining part through $M_0$. The purified geometry in this case is shown in Fig.\ref{eowRRphase2Wormhole}. It is worth noting that the RT surface of $M_1$ goes through a phase transition at late time~\cite{Balasubramanian:2020hfs}, when the purple curve $L_{M_1}$ which crosses the EOW brane replaces $M_1$ as the minimal RT surface. The same is true for $M_2$. In this phase, the reflected entropy between $R_1$ and $R_2$ corresponds to the minimal entropy associated with geodesics of $M_1 \cup M_1'$, $L_{M_1} \cup L_{M_1}'$ and $L_2 \cup L_2'$. The explicit formula of the entropy of $L_{M_1}$ can be found in~\cite{Balasubramanian:2020hfs} and we give that of $L_2$ in Appendix \ref{L1-L2-formulas}.

	\begin{figure}[htbp]
	\centering
	\includegraphics[scale=0.4]{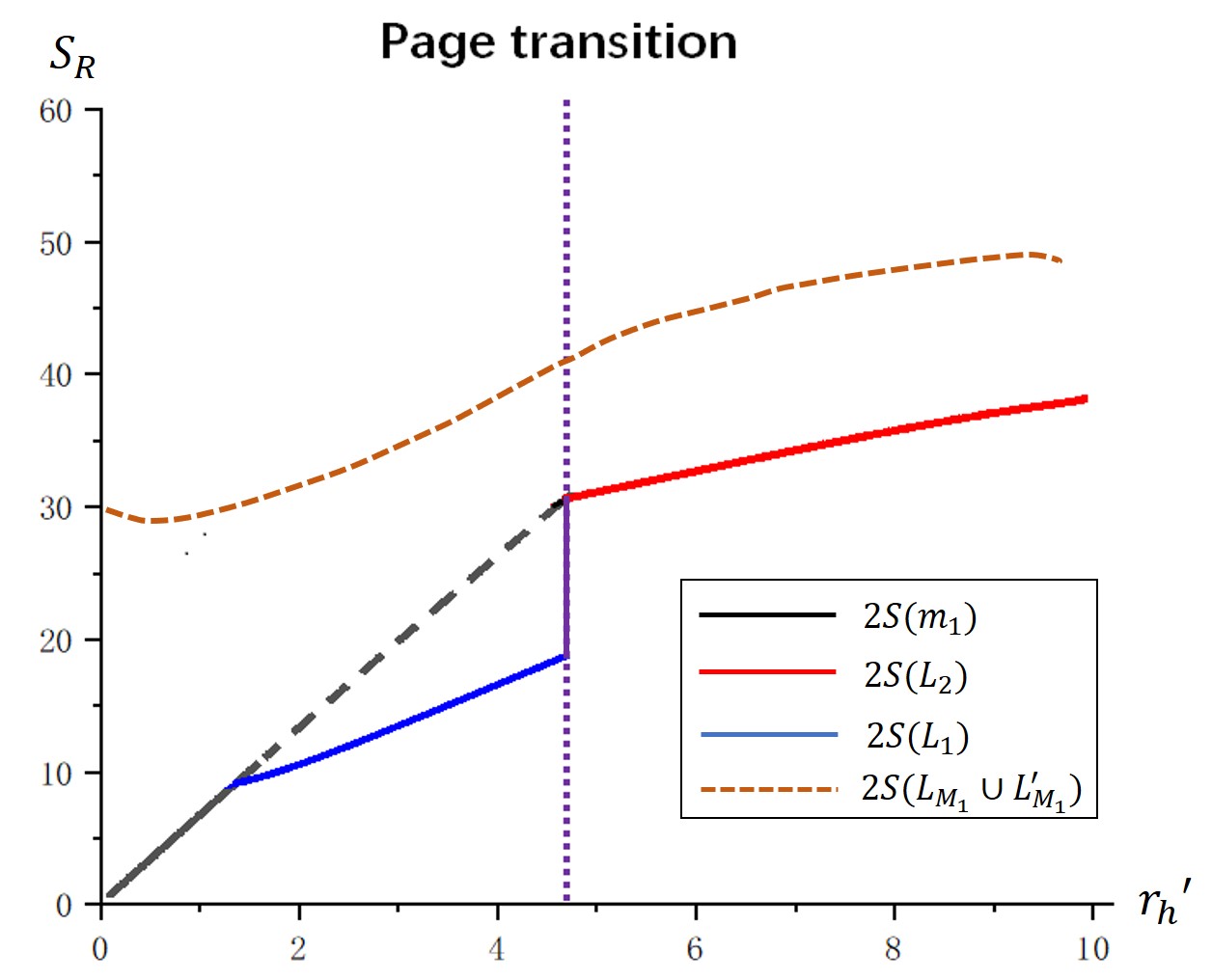}\\
	\caption{The minimum of these curves is the reflected entropy between $R_1$ and $R_2$. We can see that the reflected entropy goes through a jump at the Page transition point.}
	\label{eowRRresult}
	\end{figure}

\par
We take $r_h=10$, $l=1$, $G_N=1$, and $c'=5c$. The reflected entropy between the two parts of radiation as a function of $r_h'$ is plotted in Fig.\ref{eowRRresult}. As we can see from the figure, the reflected entropy is non-vanishing as soon as the evaporation begins, which is different from the 3-side wormhole model. The reflected entropy picks up the lowest curve at any moment of time. Notice that it goes through a jump right after the Page time ($m_3/4G_N'=m_0/4G_N$, estimated in~\cite{Balasubramanian:2020hfs}) and then turns to the red curve.

\subsection{Reflected entropy between $R_2$ and $B$}
\par
\begin{figure}[htbp]
	\centering
	\includegraphics[scale=0.4]{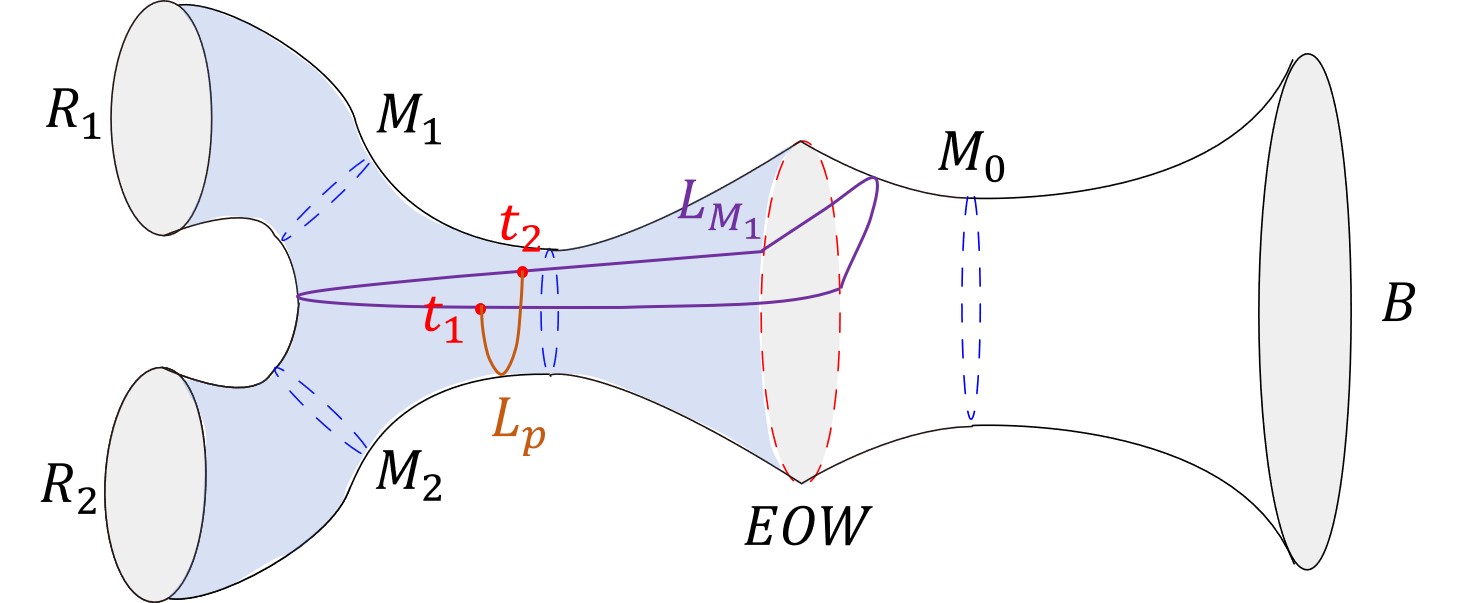}\\
	\caption{In late time of the evaporation process, the purple curve $L_{M_1}$ which crosses the EOW brane becomes the minimal RT surface homologous to $R_1$. After removing the entanglement wedge of $R_1$, the brown geodesic $L_p$ which has two intersection points on the purple curve is one possible cross section that separates $R_2$ and $B$.}
	\label{eowRBphase2}
	\end{figure}

Now we consider the reflected entropy between one part of the radiation $R_2$ and the black hole $B$. We need to remove the entanglement wedge of $R_1$ and replicate the remaining geometry to get the purified geometry. As mentioned earlier, in early time of the evaporation, the RT surface homologous to $R_1$ is $M_1$ (Fig.\ref{eowRBphase2}). But at late time after the Page time, the purple curve $L_{M_1}$ in Fig.\ref{eowRBphase2} will dominate. These two phases correspond to two different glued geometry. We will discuss the reflected entropy between $R_2$ and $B$ in these two phases separately.

\par
\begin{figure}[htbp]
	\centering
	\includegraphics[scale=0.4]{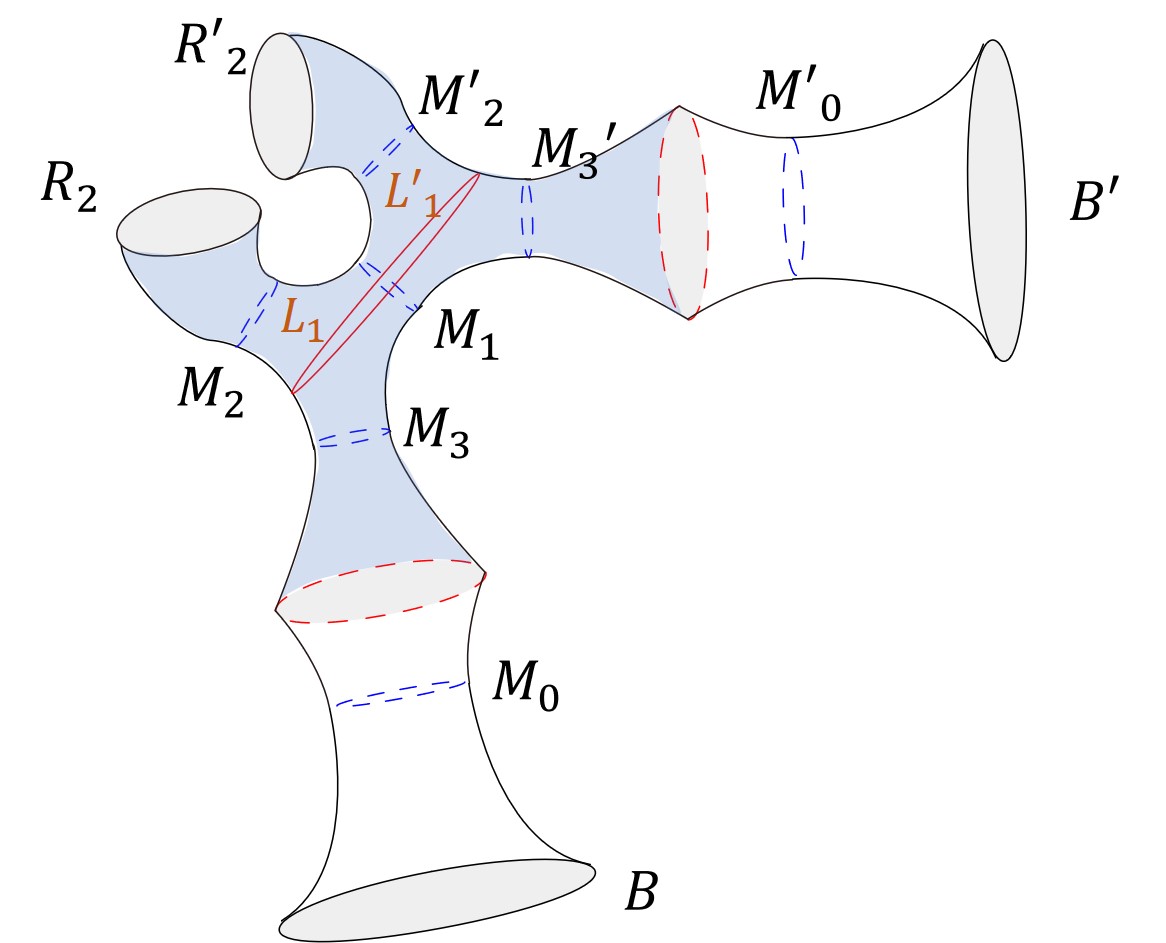}\\
	\caption{The purification geometry after tracing out $R_1$.}
	\label{eowRBphase1}
	\end{figure}

When $M_1$ is the dominant RT surface, we remove the entanglement wedge of $R_1$ and glue the two copies of the remaining geometry through $M_1$ (Fig.\ref{eowRBphase1}). The competing minimal cross sections which split $R_2\cup R_2'$ and $B\cup B'$ are $M_2 \cup M_2'$, $L_1 \cup L_1'$, $M_3 \cup M_3'$ and $M_0 \cup M_0'$. The horizon length $m_1$ is equal to $m_3$ and the length of $L_1 \cup L_1'$ can be worked out using the same formula in the previous section. 
\par
When $L_{M_1}$ becomes the dominant RT surface of $R_1$, things get a bit more complicated. We need to cut the wormhole along $L_{M_1}$ in Fig.\ref{eowRBphase2}, throw away the upper part which is connected to $R_1$, and imagine gluing two copies of the remaining lower part through $L_{M_1}$. It is hard to depict the glued geometry, but we only need to focus on one half of it, say, the lower part in Fig.\ref{eowRBphase2}. We need to find the minimal cross section that separate $R_2$ and $B$ in that part and the reflected entropy is just twice of it. There are three competing geodesics, $L_p$, $M_2$, and $M_0$ (Fig.3.7). The formula of length $L_p$ is given in Appendix \ref{Lp-formula}.

	\begin{figure}[htbp]
		\centering 
		\subfigure[]{ 
			\begin{minipage}{7cm}
			\centering 
			\includegraphics[scale=0.3]{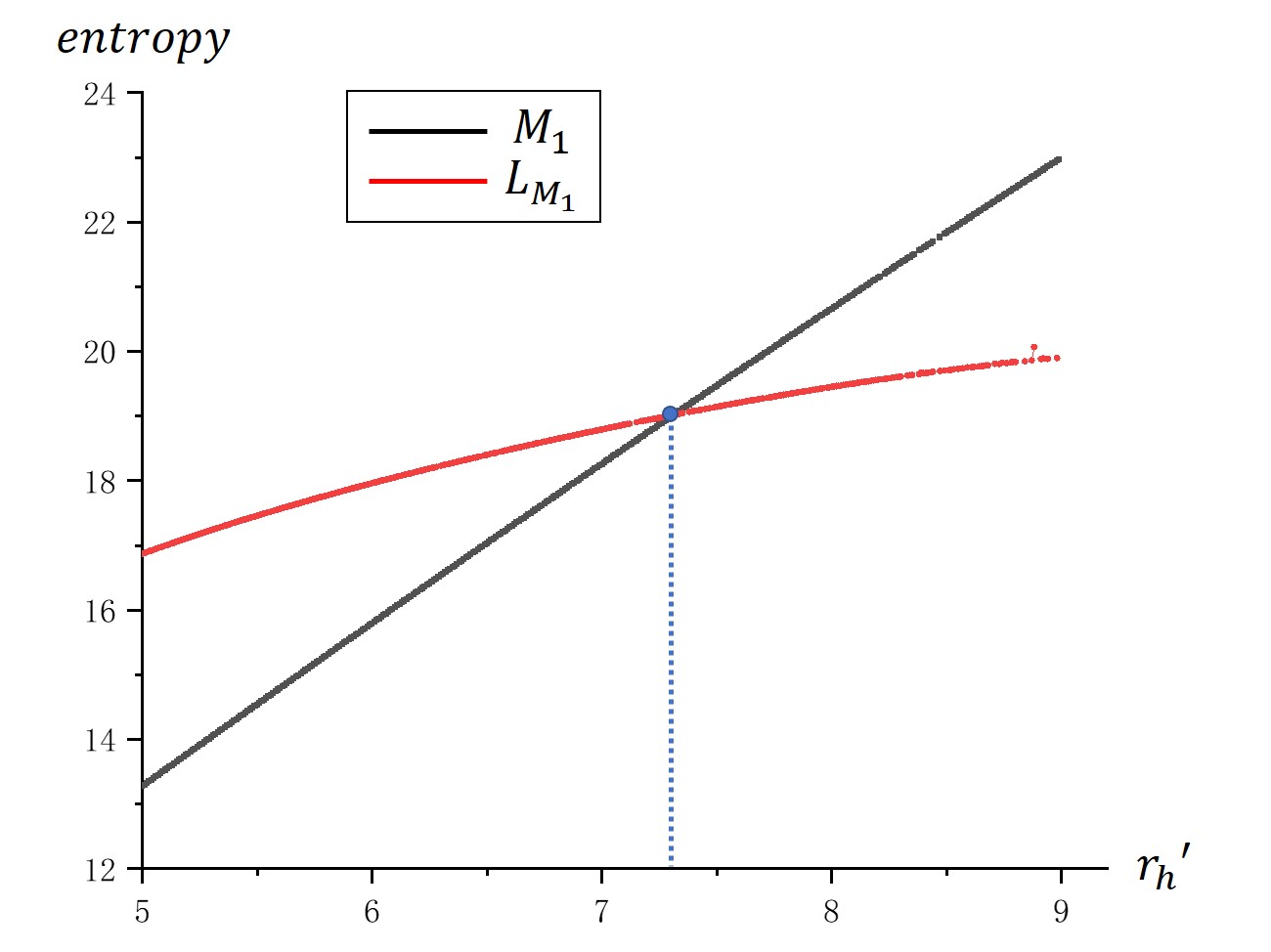} 
			\end{minipage}
		}
		\subfigure[]{ 
			\begin{minipage}{7cm}
			\centering 
			\includegraphics[scale=0.3]{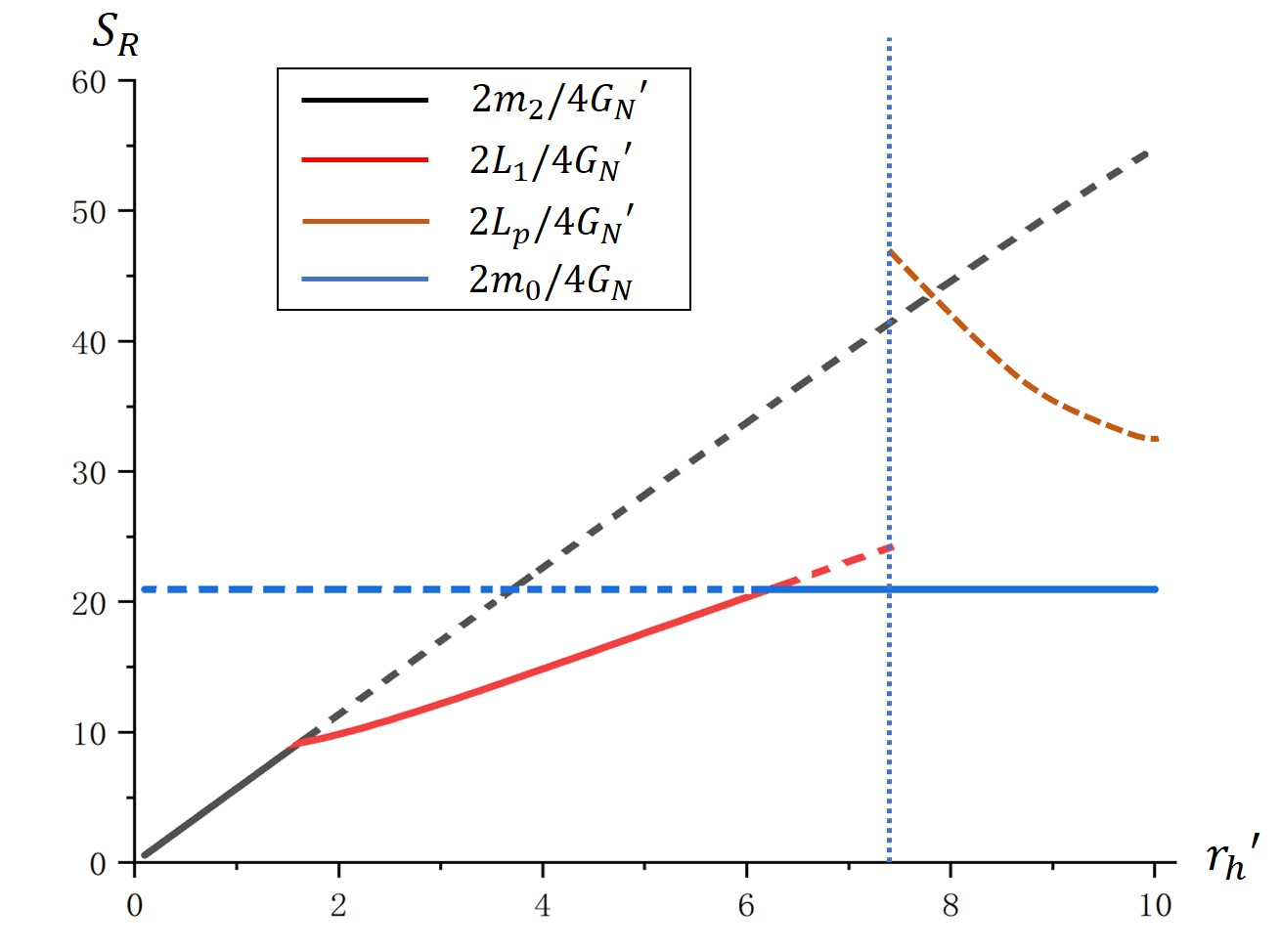} 
			\end{minipage}
		}
		\caption{(a) The entropy associated to $M_1$ and $L_{M_1}$. The RT surface of $M_1$ is one of the two curves that has smaller entropy. Note that the RT surface becomes $L_{M_1}$ when $r_h'>7.3$. (b)We take $r_h=10$, $l=1$, $G_N=1.5$ and $c'=5c$. The reflected entropy is the solid curve which saturates when $r_h' \approx 6.$} 
		\label{eowRBresult} 
	\end{figure}

\par
We plot the reflected entropy in Fig.\ref{eowRBresult} (b). The reflected entropy chooses the lowest curve. The intersection of the blue curve and the black curve is precisely the Page transition point at which $m_0/4G_N=m_3/4G_N'$ (we have $m_1=m_2=m_3$). One can see that the saturation of the reflected entropy happens later than the Page time. It is natural because the reflected entropy is bounded from above by twice of the entanglement entropy of the whole radiation, which is the black curve in Fig.\ref{eowRBresult} (b), therefore will saturate later.

\section{Quantum extremal cross section}
\begin{figure}[htbp]
	\centering
	\includegraphics[scale=0.4]{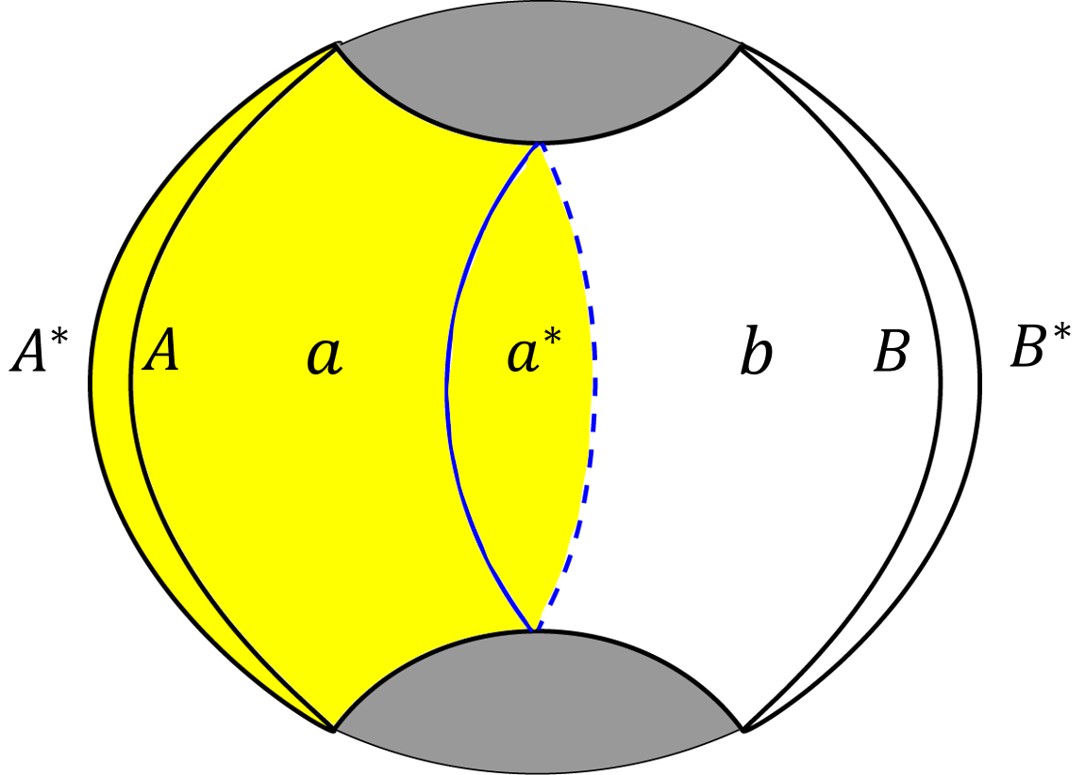}\\
	\caption{Double replica of the entanglement wedge as the bulk dual of canonical purification.}
	\label{fig1}\label{doublereplica}
	\end{figure}

Dutta and Faulkner proposed that the holographic dual of CFT reflected entropy $S_R(A:B)$ is twice the entanglement wedge cross section in the classical gravity limit of AdS/CFT~\cite{Dutta:2019gen}. The authors also conjectured the quantum corrected reflected entropy formula
\begin{equation}
S_R(A:B) = {2\langle {\cal A}[\partial a\cap\partial b]\rangle_{\tilde\rho_{ab}}\over 4G_N} + S_R^{\text{bulk}}(a:b)+{\cal O}(G_N)\label{correctedR}
\end{equation}
where the entanglement wedge of $AB$ is divided into two regions $a$,$b$ by the cross section $\partial a\cap\partial b$, and ${\cal A}$ is the area operator.\footnote{Here we focus on the static case and employ quantum extremal surface (instead of RT surface of $AB$) to define the entanglement wedge of $AB$.} $S_R^\text{bulk}(a:b)$ is the reflected entropy for the density matrix $\tilde\rho_{ab}$ of the bulk field theory. $m(AA^*)$ is the minimal surface for $AA^*$ in the double replica of the bulk entanglement wedge of $\rho_{AB}$, shown in Fig.\ref{doublereplica}.

In fact, the quantum corrected reflected entropy formula (\ref{correctedR}) can be derived from the Faulkner, Lewkowycz and Maldacena (FLM) formula of entanglement entropy~\cite{Faulkner:2013ana}. Given a bipartite density matrix $\rho_{AB}$ of a holographic CFT state, a canonical purification of $\rho_{AB}$ demands the double replica of the bulk entanglement wedge of $\rho_{AB}$ shown in Fig.\ref{doublereplica}. Now there are two boundaries $AA^*$ and $BB^*$ which support a pure state $\sqrt{\rho_{AB}}$ and the entire geometry looks like a two-side wormhole. Write the FLM formula for the entanglement entropy of one side $AA^*$ ($S_{AA^*}=S_{BB^*}$), one obtains~\footnote{Through this paper we use $S$ to denote von Neumann entropy and $S_R$ to denote reflected entropy.}
\begin{equation}
S(AA^*) = {1\over 4G_N}\langle{\cal A}[m(AA^*)]\rangle + S^{\text{bulk}}(aa^*)+{\cal O}(G_N)\ ,
\end{equation} where $aa^*$ is the entanglement wedge for $AA^*$ and $S^{\text{bulk}}(aa^*)$ is the von Neumann entropy for the bulk density matrix. The $Z_2$ symmetry ensures 
$$\langle{\cal A}[m(AA^*)]\rangle =2\langle{\cal A}[\partial a\cap\partial b]\rangle$$ and the double replica of the bulk tells that
\begin{equation}
S_R^{\text{bulk}}(a:b) = S^{\text{bulk}}(aa^*)\ .
\end{equation}
Therefore FLM of the double replica gives the quantum corrected reflected entropy formula.

Notice that FLM formula only computes the first two orders as an approximation. Engelhardt and Wall proposed that holographic entanglement entropy can be calculated exactly~\cite{Engelhardt:2014gca} in bulk Plank constant using the so called ``quantum extremal surface (QES)'' which extremizes the generalized entropy (which coincides with FLM if evaluated on the classical minimal surface).~\footnote{See~\cite{Dong:2017xht} for further discussions.} Given that reflected entropy can be realized as the entanglement entropy on canonically purified state in the level of exact density matrix, it is tempting to find a ``quantum extremal cross section (QECS)'' which can provide exact result for reflected entropy. Again we first write down the QES formula for the entanglement entropy of $S(AA^*)$
\begin{equation}
S(AA^*)= \text{ext}_Q\bigg\{{\text{Area}(Q)\over 4G_N} + S^{\text{bulk}}(aa^*)\bigg\}\ .
\end{equation} Reduced to the single replica, this becomes the extremization formula for reflected entropy
\begin{equation}
S_R(A:B)= \text{ext}_{Q'}\bigg\{{2 \text{Area}(Q'=\partial a\cap\partial b)\over 4G_N} + S_R^{\text{bulk}}(a:b)\bigg\}\ ,
\end{equation} where the quantum extremal cross section is denoted by $Q'$. This is our main proposal in this section.

Recently it has been proposed that QES formula can compute the fine-grained entropy not only for subregions of holographic CFT states but also for general gravitational systems including black holes and quantum systems coupled with gravity. See~\cite{Almheiri:2020cfm} for a recent discussion on the fine-grained gravitational entropy. Specifically, the fine-grained entropy of AdS black hole surround by matter is given by the generalized entropy of QES,
\begin{equation}
S_B = \text{ext}_Q\bigg\{{\text{Area}(Q)\over 4G_N} + S(\tilde\rho_B) \bigg\}\ ,\label{bhQES}
\end{equation}
where $Q$ is the quantum extremal surface, and $B$ is the region between $Q$ and AdS boundary. For a quantum system coupled to gravity, such as the CFT bath in the recent 2d JT gravity+CFT model of black hole evaporation, the von Neumann entropy of bath CFT is given by
\begin{equation}
S(\rho_R) = \text{ext}_I\bigg\{{\text{Area}(\partial I=Q)\over 4G_N} + S(\tilde\rho_{R\cup I}) \bigg\}\ .\label{islandQES}
\end{equation}
Importantly, an island contribution has to be included, which can be derived by the gravitational path integral calculation of the von Neumann entropy~\cite{Almheiri:2019qdq,Lewkowycz:2013nqa}. If there is more than one extremum, then $Q$ is the surface with minimal entropy. Notice that trivial island is always an extremal solution for (\ref{islandQES}), where
\begin{equation}
S(\rho_R) = S(\tilde\rho_R)\ ,
\end{equation}
therefore the island solution is preferred only if the entropy with island is less than the one without island.

The formula (\ref{bhQES}) can be considered as the black hole version of the original QES and (\ref{islandQES}) can be considered as the radiation version of QES. Since reflected entropy can always be realized as the entanglement entropy in a canonically purified state, it is tempting to find similar generalizations of QECS for reflected entropy. In the following section we will derive some generalizations of QECS by looking into the two-dimensional eternal black hole + 2d CFT model of black hole evaporation. The reason is that this model has a left/right $Z_2$ symmetry and the right half can be considered as the canonical purification of the left. The eternal black hole + 2d CFT model provides a natural framework to establish the generalizations of QECS.

\section{Reflected entropy in 2d Eternal black hole + CFT model}\label{sec4}
In this section we consider a model where a two-side eternal black hole with Jackiw-Teitelboim gravity is coupled to a bath CFT. The model was analyzed in great detail in~\cite{Almheiri:2019qdq} for the purpose of resolving the black hole information paradox.
\subsection{Review of the model}
In this model, a AdS$_2$ region with JT gravity and Minkowski spacetime are glued together. In addition, we have a large central charge CFT living in both the AdS$_2$ and the flat spacetime and one can impose a transparent boundary condition. The action is given by
\begin{equation}
I_{\text{total}} = -{S_0\over 4\pi} \biggr [\int_\Sigma R + \int_{\partial\Sigma} 2K \biggr] - \int_\Sigma (R+2){\phi\over 4\pi} - {\phi_b\over 4\pi}\int_{\partial\Sigma} 2K+ S_{\text{CFT}}\ ,
\end{equation}
where the spacetime is dynamical in $\Sigma$ but rigid in the exterior region. We will set $4G_N=1$ and the area term of the entropies will be given by $S_0+\phi$.

	\begin{figure}[htbp]
	\centering
	\includegraphics[scale=0.4]{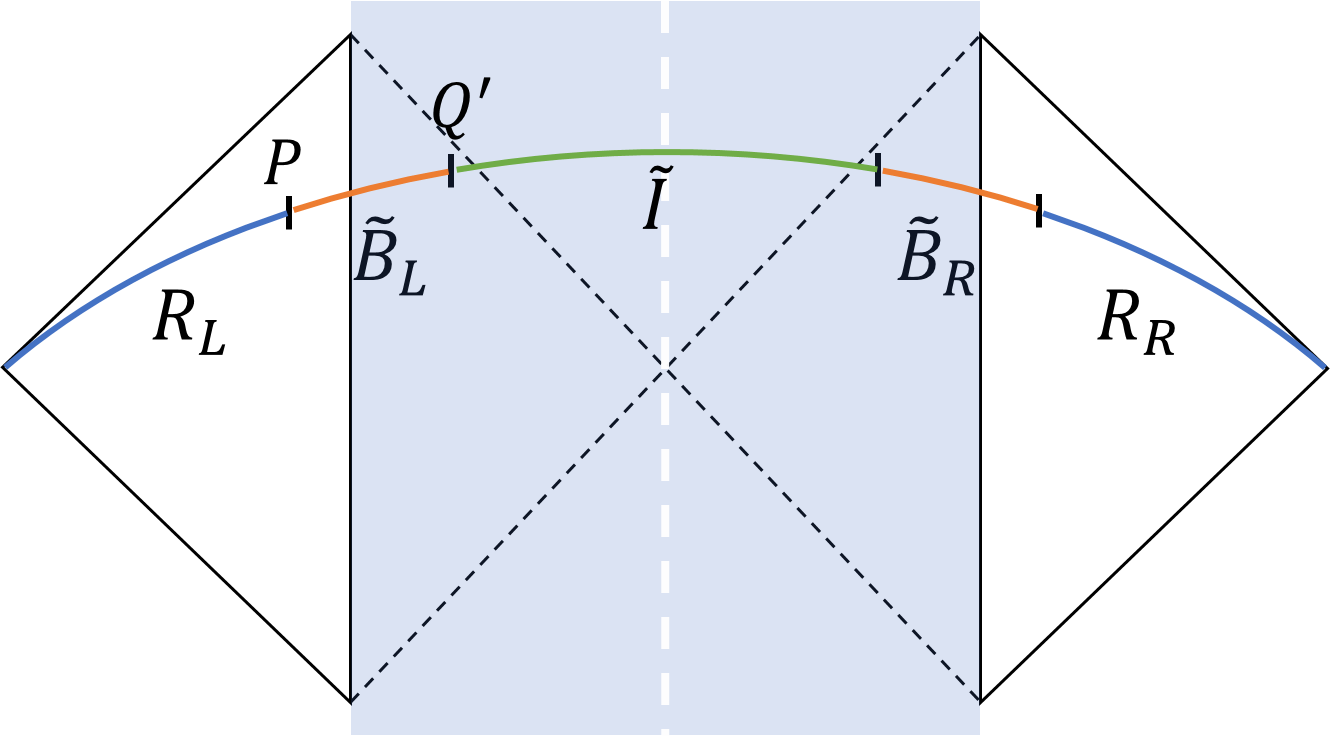}\\
	\caption{Radiation and black hole on one side and the reflection on the other side. The $(\sigma, t)$ coordinates are $P:(b,-t+\pi i), Q':(-a,-t+\pi i)$.}
	\label{jtAC}
	\end{figure}

Now we focus on the eternal black hole solution with the dilaton (vacuum solution in the bulk). As shown in Fig.\ref{jtAC} in Lorentzian signature, the eternal black hole lives in the gravitational region (shaded) while the non-gravitational region (non-shaded) plays the role of radiation reservoir. Their metrics are
\begin{equation}
ds^2_{\text{gravity}}=\frac{4\pi^2}{\beta^2}\frac{dyd\bar{y}}{\sinh^2\frac{\pi(y+\bar{y})}{\beta}},\ \ ds^2_{\text{non-gravity}}=\frac{dyd\bar{y}}{\epsilon^2}\ ,
\end{equation}
where the complex coordinates in Euclidean signature is $y=\sigma+i\tau$ and $\bar{y}=\sigma-i\tau$ and the inverse temperature is denoted by $\beta$. The boundary of the gravitational region is located at $\sigma=-\epsilon$. Lorentzian time is related by $\tau=-i t$ . Upon the transformation $w=e^{\frac{2\pi y}{\beta}}$, the metrics become
\begin{equation}
\label{metricw}
\begin{split}
ds^2_{\text{gravity}}=\frac{4dw d\bar{w}}{(1-w\bar{w})^2},\ \ ds^2_{\text{non-gravity}}=\frac{\beta^2 dwd\bar{w}}{4\pi^2\epsilon^2w\bar{w}}\ ,
\end{split}
\end{equation}
from which conformal factors in a general form d$s^2=\Omega^{-2}\text{d}w\text{d}\bar w$ can be read directly,
\begin{equation}
\Omega_{\text{gravity}}=\frac{1-w\bar{w}}{2},\ \ \Omega_{\text{non-gravity}}=\frac{2\pi \epsilon}{\beta}\sqrt{w\bar{w}}\ .
\end{equation}
The dilaton solution only defined in the gravitational region is given by
\begin{equation}
\phi = -\frac{2\pi}{\beta}{\phi_r\over \tanh \frac{2\pi \sigma}{\beta}}\ ,
\end{equation} with $\phi = \phi_r/\epsilon$ at the boundary.
A time slice can be considered as a pure quantum state, made up of black hole $B$, radiation $R$ and island $I$. We can compute the fine grained entropies by (\ref{bhQES}) and (\ref{islandQES}). In those formulas, the area term at point $(\sigma,t)$ is the value of the dilaton $\phi$ plus the genus-counting parameter $S_0$ of JT gravity.
\subsection{A formula of reflected entropy}
As shown in Fig.\ref{jtAC}, the left quantum system is divided into two, $R_L$ and $B_L$.
The von Neumann entropy of $B_L$ and $R_L$ are given by the QES formula (\ref{bhQES}) and (\ref{islandQES})
\begin{equation}
S(B_L) = \text{min}\,\,\text{ext}_Q\bigg\{{A(Q)\over 4G_N} + S(\tilde\rho_{B_L}) \bigg\}\ ,\label{QES0}
\end{equation}
\begin{equation}
S(R_L) = \text{min}\,\,\text{ext}_Q\bigg\{{A(Q=\partial I_L)\over 4G_N} + S(\tilde\rho_{R_L\cup I_L}) \bigg\}\ .\label{QES1}
\end{equation}
The reflected entropy of $R_L$ and $B_L$ can be measured as the entanglement entropy between $R=R_L\cup R_R$ and $B=B_L\cup B_R$ by treating the right part as the canonical purification of the left part.~\footnote{It is noted in~\cite{Dutta:2019gen} that the two-side eternal black hole can be treated as a canonical purification of the single side CFT state.} Therefore
\begin{equation}
S_R(R_L:B_L) = S(R)= \text{min}\,\,\text{ext}_I\bigg\{{A(\partial I)\over 4G_N} + S(\tilde\rho_{R\cup I}) \bigg\}\ ,\label{QESofR}
\end{equation}
which is the formula reproducing the Page curve of the total radiation, respecting the unitarity, during black hole evaporation.
Notice that (\ref{QESofR}) preserves a left/right $Z_2$ symmetry. Reduced to the left side, the formula becomes
\begin{equation}
S_R(R_L:B_L) = \text{min}\,\,\text{ext}_{Q'}\bigg\{{2A(Q'=\partial \tilde I_L\cap \partial \tilde B_L)\over 4G_N} + S_R(\tilde\rho_{R_L\cup \tilde I_L}:\tilde\rho_{\tilde B_L}) \bigg\}\ ,\label{QECS1}
\end{equation} 
where $Q'$ is the cross section. Notice that $\tilde I_L\cup \tilde B_L$ is the whole left bulk,~\footnote{It is verified by QES calculation in Appendix \ref{bounds} that the whole left quantum system corresponds to the whole left bulk precisely.} but in general $\tilde I_L$ and $\tilde B_L$ are not $I_L$ and $B_L$ in (\ref{QES0})(\ref{QES1}). Eq.(\ref{QECS1}) is our main result in this section. It can be considered as a general formula to compute the reflected entropy between (part of) radiation and black hole. Notice that generally some part of radiation together with the black hole is not a pure state.

(\ref{QECS1}) can also be considered as the generalized quantum extremal cross section formula. The interesting thing is that, in the generalized QECS, the cross section is associated to the island. When $R_L=0$ or $B_L=0$, it vanishes because both the area term and the second term vanish. Now we give another test of this formula by computing the reflected entropy for a pure state, $S_R(R_L\cup R_R:B_L\cup B_R)$. Following the formula of (\ref{QECS1})
\begin{equation}
S_R(R:B) = \text{min}\,\,\text{ext}_{Q'}\bigg\{{2A(Q'=\partial \tilde I\cap \partial \tilde B)\over 4G_N} + S_R(\tilde\rho_{R\cup \tilde I}:\tilde\rho_{\tilde B}) \bigg\}\ .
\end{equation} 
Since the state on the whole slice is pure we have $S_R(\tilde\rho_{R\cup \tilde I}:\tilde\rho_{\tilde B})=2S(\tilde\rho_{R\cup \tilde I})$. Also $\partial \tilde I\cap \partial \tilde B=\partial \tilde I$ because the two boundaries are identical. Therefore we get twice of the island formula of radiation in the right hand side, which is consistent with the information theoretical relation $S_R(R:B)=2S(R)$ for pure state of $RB$.

In the rest of this section, we illustrate several examples of computing reflected entropy.
\subsection{Radiation and Black hole}\label{sec43}
We follow (\ref{QECS1}) to compute the reflected entropy of the left radiation and the left black hole. From now  on we set the inverse temperature $\beta=2\pi$.
	
	\begin{figure}[htbp]
         \centering
	\includegraphics[scale=0.4]{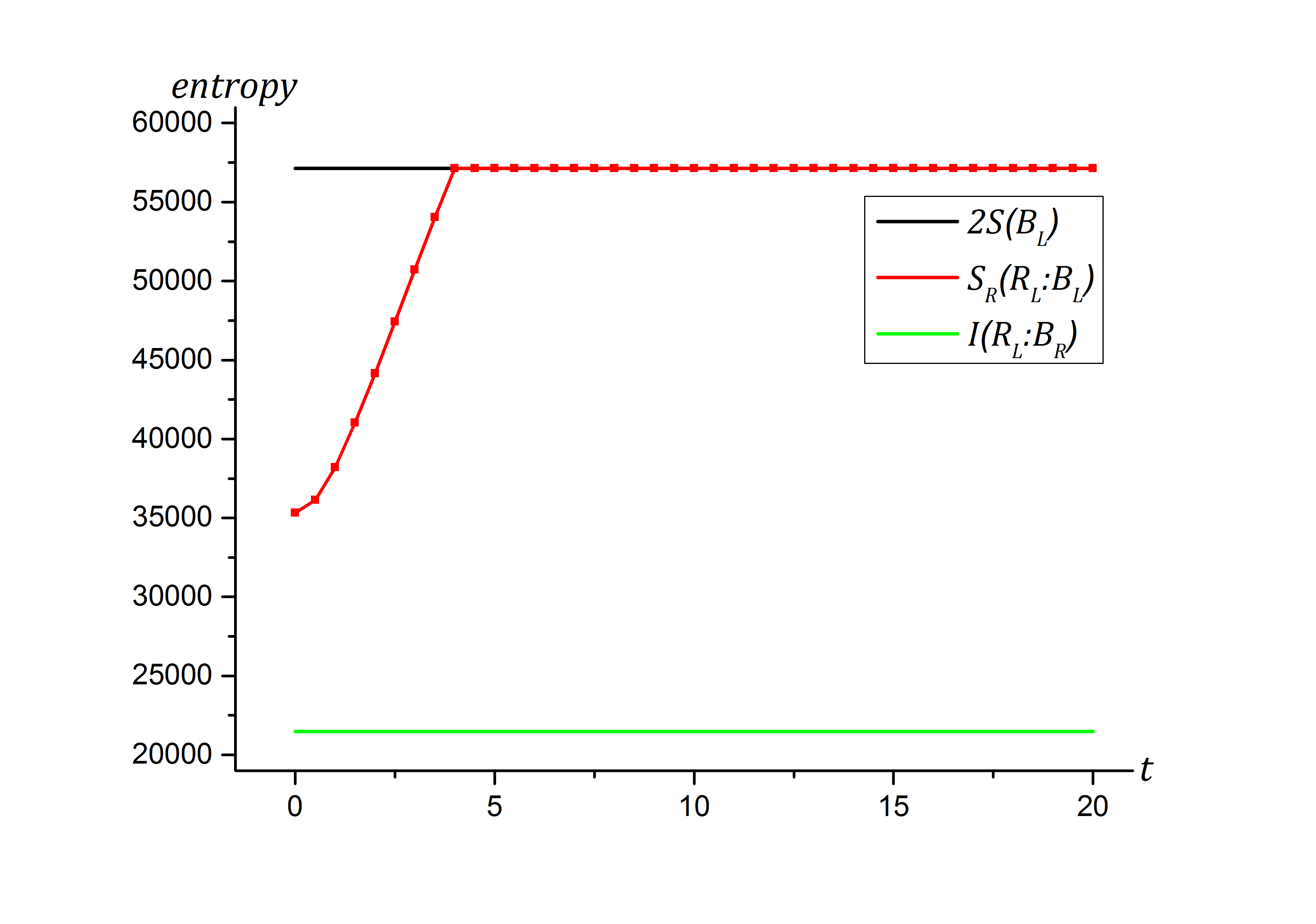}\\
	\caption{Reflected entropy $S_R(R_L:B_L)$ as a function of time $t$. We pick $b=0.01, \phi_r=100, S_0=c=20000$ and $\epsilon_{UV}=0.01$. It is calculated that the upper bound $2S(B_L)=57146.50$ and the lower bound $I(R_L:B_L)=21479.99$.}
	\label{jtSRAC}
	\end{figure}

As shown in Fig.\ref{jtAC}, $R_L$ joins with $B_L$ at the point $P:(b,-t+\pi i)$. When ${\tilde I_L}$ is finite, ${\tilde B_L}$ joins with ${\tilde I_L}$ at the point $Q':(-a,-t+\pi i)$. So the cross section term in (\ref{QECS1}) is
\begin{equation}
2(S_0+\frac{\phi_r}{\tanh a})\ .
\end{equation}
The second term in (\ref{QECS1}) is just the von Neumann entropy of ${\tilde B_L}\cup{\tilde B_R}$, whose formula is given in free fermion theory by~\cite{Almheiri:2019qdq}
\begin{equation}
S({\tilde \rho_{{\tilde B_L}\cup{\tilde B_R}}})=\frac{c}{3}\ln \frac{2\cosh^2 t (\cosh (a+b)-1)}{\sinh a \cosh (\frac{a+b-2t}{2}) \cosh (\frac{a+b+2t}{2})}-\frac{c}{3}\ln \epsilon_{UV}\ .
\end{equation}
Extremizing two terms together gives the QES equation for $a$, namely
\begin{equation}
\label{QESa}
\begin{split}
&\frac{c}{3}(\frac{\sinh (a+b)}{\cosh (a+b)-1}-\coth a-\frac{1}{2}\tanh \frac{a+b-2t}{2}-\frac{1}{2}\tanh \frac{a+b+2t}{2})\\
=&2\frac{\phi_r}{\sinh^2 a} \ .
\end{split}
\end{equation}

When ${\tilde I_L}=\emptyset$, we will have $Q'={\tilde B_L}\cap {\tilde I_L}=\emptyset$, which means that the cross section term vanishes. Therefore, 
\begin{equation}
\label{rbni}
S_R(R_L:B_L)=S({\tilde \rho_{{\tilde B_L}\cup{\tilde B_R}}})=\frac{c}{3}\ln (2\cosh t)-\frac{c}{3}\ln \epsilon_{UV} \ .
 \end{equation}
 
 The final result of $S_R(R_L:B_L)$ is given by the minimum of the above two.
 
In Fig.\ref{jtSRAC} we plot $S_R(R_L:B_L)$ with the upper bound $2S(B_L)$ and the lower bound $I(R_L:B_L)$ which are independent of time (see Appendix~\ref{bounds} for the calculation of the bounds). Notice that $S_R(R_L:B_L)$ is just the Page curve shown in~\cite{Almheiri:2019qdq}, which increases at early time and saturates at late time.
\subsection{Black hole and Black hole}\label{sec42}
	\begin{figure}[htbp]
	\centering
	\includegraphics[scale=0.4]{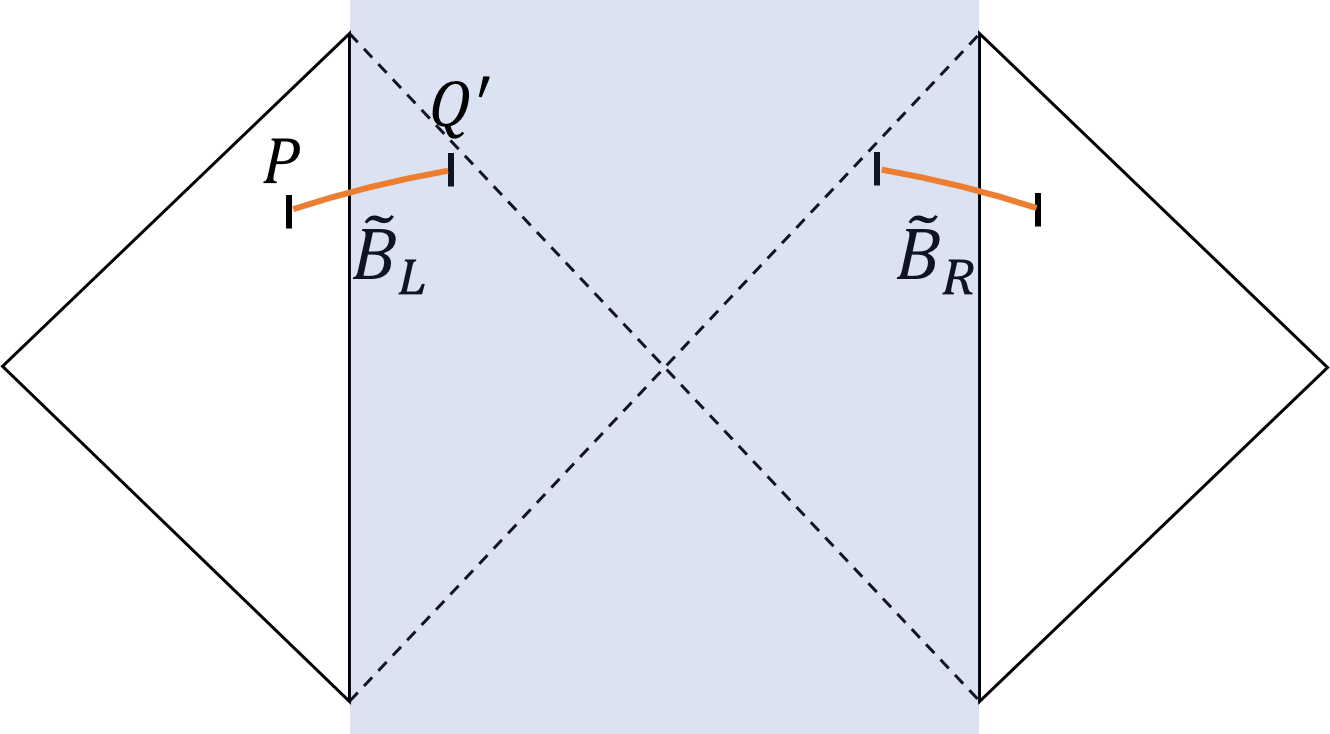}\\
	\caption{$\tilde B_L$ and $\tilde B_R$ are two disjoint intervals in the presence of an island.}
	\label{jtCD}
	\end{figure}
It is easy to generalize (\ref{QECS1}) to the reflected entropy between the left black hole and the right black hole
\begin{equation}
S_R(B_L:B_R) = \text{min}\,\,\text{ext}_{Q'}\bigg\{{2A(Q'=\partial \tilde B_L\cap \partial \tilde B_R)\over 4G_N} + S_R(\tilde\rho_{\tilde B_L}:\tilde\rho_{\tilde B_R}) \bigg\}\ .
\end{equation} Notice that $\tilde B_L\cup\tilde B_R$ is the bulk region of two side black holes, but in general they are not $B_L$ and $B_R$.

At early time, the radiation region $R$ has no island. Therefore $\tilde B_L$ intersects with $\tilde B_R$ at the middle point $w_0=0$. We can compute $S_R(\tilde\rho_{\tilde B_L}:\tilde\rho_{\tilde B_R})$ by three-point correlation functions of twist operators $\sigma_i$,
\begin{equation}
\label{tpcf}
S_R(\tilde\rho_{\tilde B_L}:\tilde\rho_{\tilde B_R})=\lim_{\pmb{n} \to 1}\frac{1}{1-\pmb{n}}\ln \frac{\prod_i \Omega_i^{2h_i}\langle \sigma_{g_A}(-e^{b-t})\sigma_{g_B^{-1}}(e^{b+t})\sigma_{g_A^{-1}g_B}(0)\rangle_{CFT^{\otimes m\pmb{n}}}}{(\prod_i \Omega_i^{2h_i(\pmb{n}=1)}\langle \sigma_{g_A}(-e^{b-t})\sigma_{g_B^{-1}}(e^{b+t})\sigma_{g_A^{-1}g_B}(0)\rangle_{CFT^{\otimes m}})^{\pmb{n}}}\ ,
\end{equation} 
where $h_i$ is the conformal dimension of $\sigma_i$ and $\Omega_i$ is the associated conformal factor. The twist operator approach to compute reflected entropy can be found in~\cite{Dutta:2019gen}. The conformal dimensions are given by
\begin{equation}
\label{cdh}
h_{g_A}=h_{g_B^{-1}}=\frac{c\pmb{n}}{24}(m-\frac{1}{m}),\quad h_{g_A^{-1}g_B}=\frac{c}{12}(\pmb{n}-\frac{1}{\pmb{n}})\ ,
\end{equation} 
and we employ eq.(C.9) in~\cite{Dutta:2019gen}
\begin{equation}
\label{3pcf}
\langle \sigma_{g_A}(-e^{b-t})\sigma_{g_B^{-1}}(e^{b+t})\sigma_{g_A^{-1}g_B}(0)\rangle_{CFT^{\otimes m\pmb{n}}}=(2m|e^{b+t}||e^{b-t}|)^{-4h_{\pmb{n}}}|e^{b-t}+e^{b+t}|^{-4\pmb{n}h_m+4h_{\pmb{n}}}\ ,
\end{equation} 
where $h_{\pmb{n}}=\frac{c}{24}(\pmb{n}-\frac{1}{\pmb{n}}),h_{m}=\frac{c}{24}(m-\frac{1}{m})$. By inserting (\ref{cdh}) and (\ref{3pcf}) into (\ref{tpcf}), plus the area term we get the total reflected entropy
\begin{equation}
S_R(B_L:B_R)=2S_0+2\phi_r +\frac{c}{3}(b-\ln \cosh t+\ln 2)\ .
\end{equation} 

At late time, the radiation $R$ has an island. The black hole $B$ is then divided into two disjoint intervals $\tilde B_L$ and $\tilde B_R$. As shown in Fig.\ref{jtCD}, the cross section term vanishes. Therefore
\begin{equation}
S_R(B_L:B_R) = S_R(\tilde\rho_{\tilde B_L}:\tilde\rho_{\tilde B_R})\ .
\end{equation} 
Since the cross ratio $\eta\equiv \frac{(e^b-e^{-a})^2}{(e^{-a+t}+e^{b-t})(e^{b+t}+e^{-a-t})}$ goes to 0 when $t$ is large, we can use the approximate formula of reflected entropy in free fermion theory~\cite{Bueno:2020vnx}
\begin{equation}
S_R(\tilde\rho_{\tilde B_L}:\tilde\rho_{\tilde B_R})\sim c(-0.15\eta \ln \eta+0.67\eta)\ .
\end{equation}
Notice that in this case the conformal factors $\Omega_i$ are cancelled upon normalization as $h_{g_A}$ and $h_{g_B^{-1}}$ in (\ref{tpcf}) because for these operators we have $h_i=\pmb{n}h_i(\pmb{n}=1)$ (see (\ref{cdh})).

The upper bound $2S(B_L)=2S(B_R)$ has been calculated in (\ref{ebl}). And $S(B_L\cup B_R)$ is equal to $S_R(R_L:B_L)$ as calculated in Section \ref{sec43}, since they share the same formula according to the definition of reflected entropy. Therefore, the lower bound $I(B_L:B_R)=2S(B_L)-S(B_L\cup B_R)$ is also known.

	\begin{figure}[htbp]
	\centering
	\includegraphics[scale=0.4]{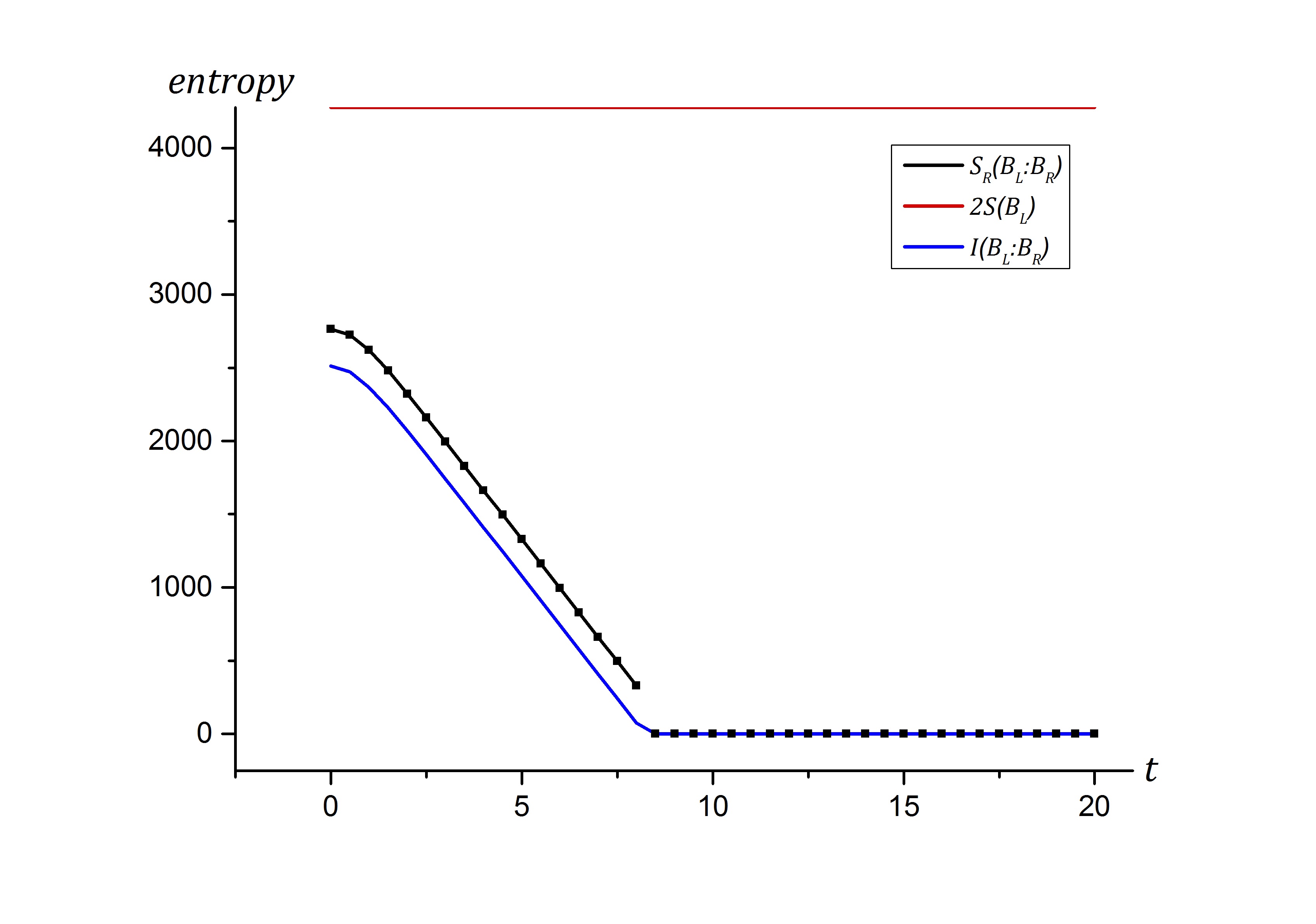}\\
	\caption{$S_R(B_L:B_R)$ with the upper bound $2S(B_L)$ and the lower bound $I(R_L:B_L)$, with respect to $t$. We pick $b=1,\phi_r=100, S_0=c=1000$ and $\epsilon_{UV}=0.01$. Notice that the numerical data shows that at late time $S_R(B_L:B_R)$ is very closed (but still greater) to $I(B_L:B_L)$.}
	\label{jtSRCD}
	\end{figure}
	
We plot $S_R(B_L:B_R)$ as well as its bounds in Fig.\ref{jtSRCD}. It shows that the correlation between the left black hole and the right black hole decreases and goes to zero at late time.

\subsection{Radiation and Radiation}\label{sec41}
	\begin{figure}[htbp]
	\centering
	\includegraphics[scale=0.4]{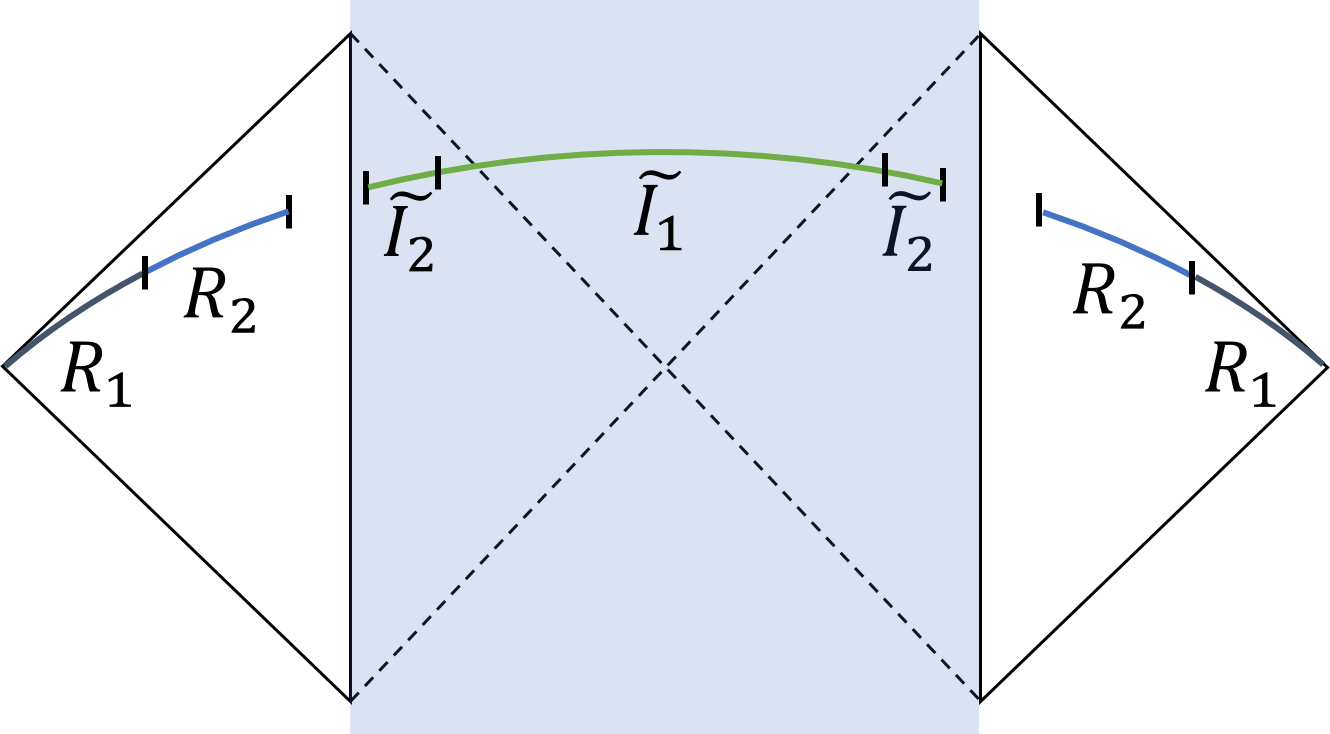}\\
	\caption{Two subsystems $R_1$ and $R_2$ in the radiation. The island $I$ of $R=R_1\cup R_2$ is divided into two parts $\tilde I_1$ and $\tilde I_2$.}
	\label{jtA'B'1}
	\end{figure}
	
It is easy to generalize (\ref{QECS1}) to the reflected entropy between radiation and radiation,
\begin{equation}
S_R(R_1:R_2) = \text{min}\,\,\text{ext}_{Q'}\bigg\{{2A(Q'=\partial \tilde I_1\cap \partial \tilde I_2)\over 4G_N} + S_R(\tilde\rho_{R_1\cup\tilde I_1}:\tilde\rho_{R_2\cup\tilde I_2}) \bigg\}\ .\label{QECS3}
\end{equation}
Notice that $\tilde I_1\cup\tilde I_2$ (see Fig.\ref{jtA'B'1}) is the whole island of $R_1\cup R_2$ but in general they are not $I_1$ and $I_2$.
	\begin{figure}[htbp]
	\centering
	\includegraphics[scale=0.4]{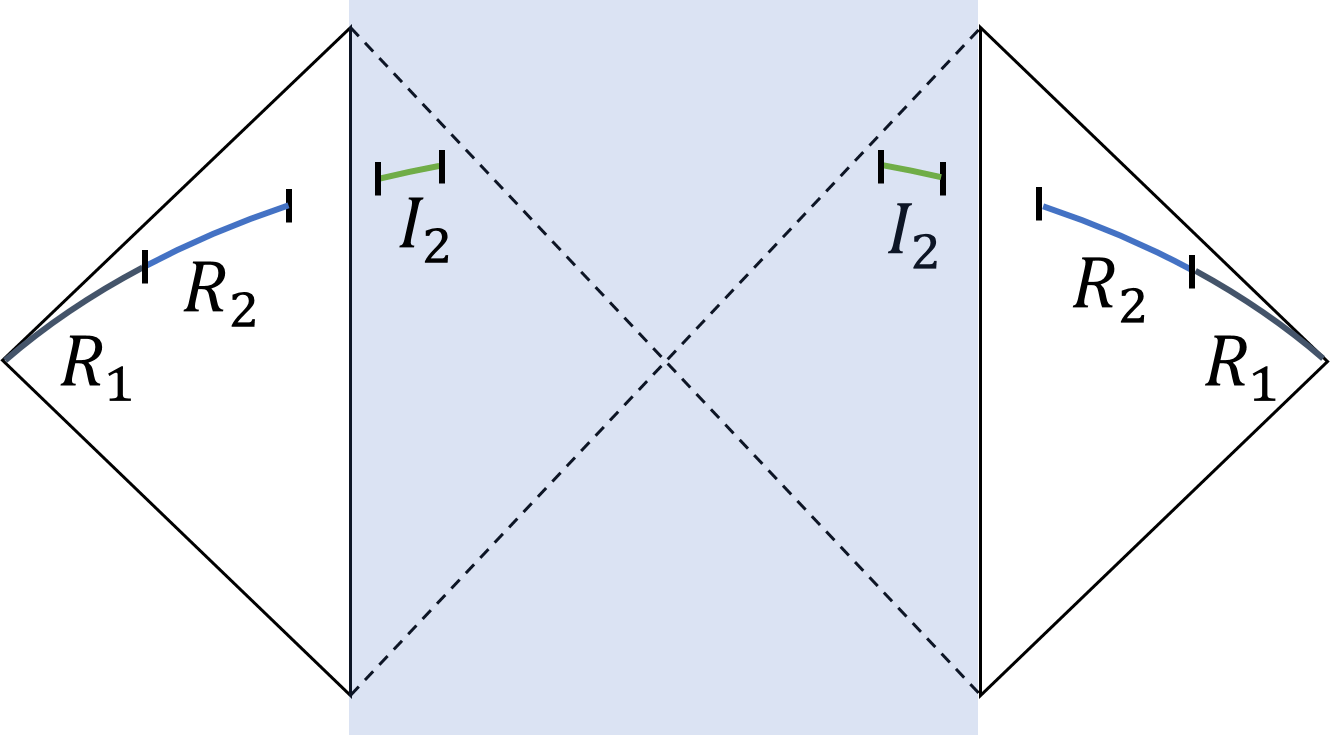}\\
	\caption{Two subsystems $R_1$ and $R_2$ in the radiation. The coordinates (from left to right) are $P_1':(b_2,-t+\pi i), P_2':(b_1,-t+\pi i), P_3':(-a_1,-t+\pi i), P_4':(-a_2,-t+\pi i),P_5':(-a_2,t), P_6':(-a_1,t), P_7':(b_1,t)$ and $P_{8}':(b_2,t)$. Here the island of $R_2$ is made up of two disjoint intervals.}
	\label{jtA'B'}
	\end{figure}

	\begin{figure}[htbp]
	\centering
	\includegraphics[scale=0.5]{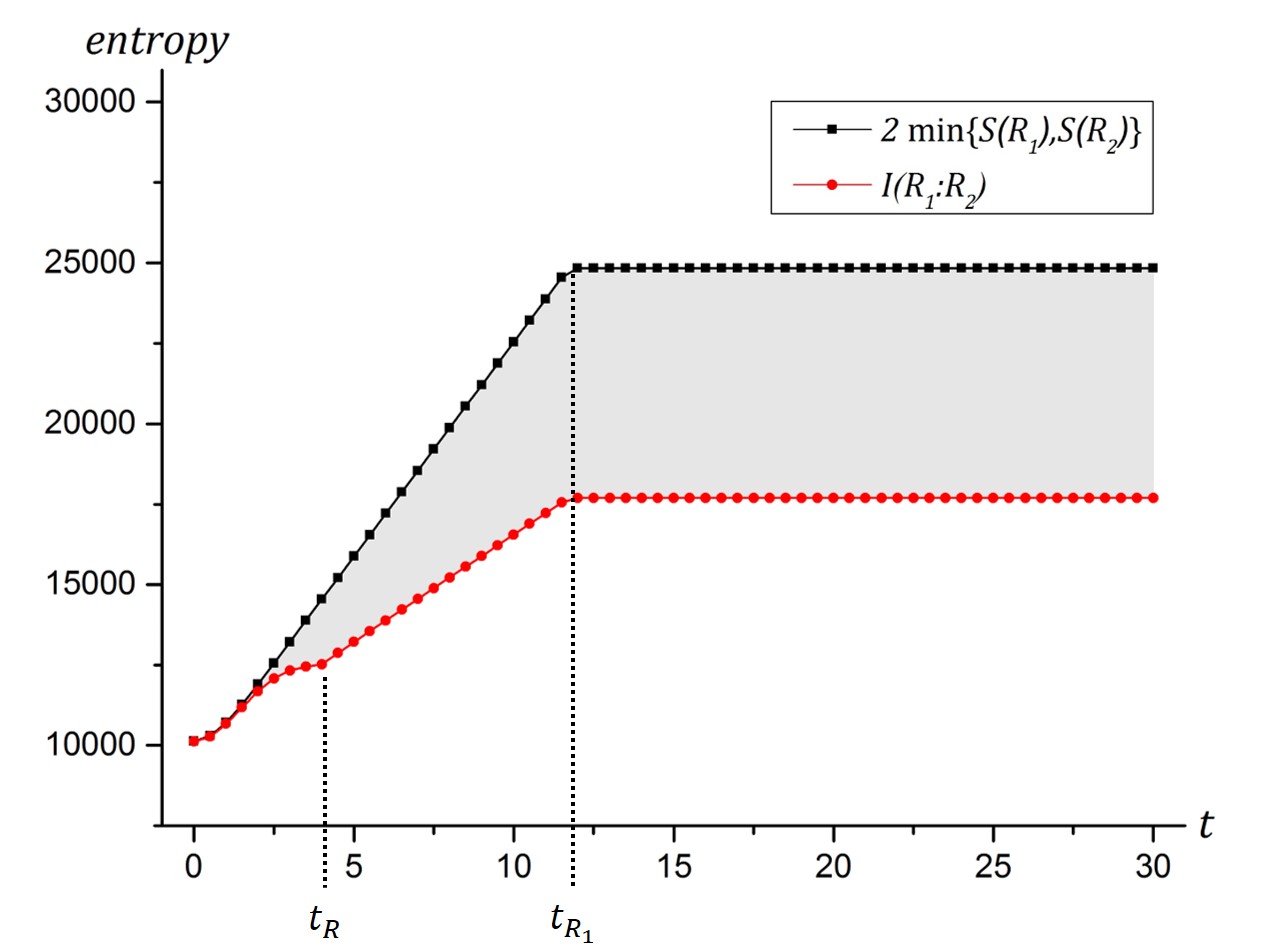}\\
	\caption{The upper bound $2\min \{S(R_1),S(R_2)\}$ and the lower bound $I(R_1:R_2)$ of reflected entropy between $R_1$ and $R_2$. We pick $b_1=0.01,b_2=5, \phi_r=10,S_0=c=2000$ and $\epsilon_{UV}=0.001$. Typical time scales are denoted in the figure: $t_{R_1},t_{R}$ denote the Page time for $R_1$ and $R_1\cup R_2$ while $R_2$ does not have an island for these parameters.}
	\label{jtSRA'B'}
	\end{figure}
Now we consider two radiation subsystems as shown in Fig.\ref{jtA'B'}. Due to the technical difficulty to compute the second term in (\ref{QECS3}), we instead only give the bounds of $S_R(R_1:R_2)$. The calculations of $S(R_1\cup R_2)$ and $S(R_1)$ are similar to that of $S_R(R_L:B_L)$ shown in Section \ref{sec43}. As for the entropy $S(R_2)$, we employ the formula for multi-interval cases in free fermion theory~\cite{Casini:2005rm}
\begin{equation}
\begin{split}
S^{\text{no island}}(R_2)=\frac{c}{3}\ln \frac{8\sinh^2(\frac{b_1-b_2}{2})\cosh^2t}{\cosh (b_1-b_2)+\cosh (2t)}-\frac{2c}{3}\ln \epsilon_{UV}\ ,
\end{split}
\end{equation}
\begin{equation}
\label{SR2}
\begin{split}
S^{\text{island}}(R_2)&=4S_0+2\frac{\phi_r}{\tanh a_1}+2\frac{\phi_r}{\tanh a_2}\\
+&\frac{c}{6}\ln \frac{\prod_{i\in \{ 1,3,5,7\},j\in \{2,4,6,8\} }|w_i'-w_j'|^2}{\prod_{i,j\in \{ 1,3,5,7\} ,i<j}|w_i'-w_j'|^2\prod_{i,j\in \{ 2,4,6,8\} ,i<j}|w_i'-w_j'|^2\prod_{i}\Omega_i'}\ ,
\end{split}
\end{equation}
where $w_i'$ denotes the coordinate of the point $P_i'$ and $\Omega_i'$ denotes its conformal factors. Note that (\ref{SR2}) should be extremized with respect to $P_3',P_4',P_5',P_6'$, which gives two equations for $a_1$ and $a_2$.
Finally,
\begin{equation}
\begin{split}
S(R_2)=\min \{S^{\text{island}}(R_2),S^{\text{no island}}(R_2)\} \ .
\end{split}
\end{equation}

We plot the bounds of $S_R(R_1:R_2)$ in Fig.\ref{jtSRA'B'}. The shaded region in Fig.\ref{jtSRA'B'} implies that reflected entropy between radiation and radiation increases at early time and saturates at late time.

\section{Conclusion and Discussion}\label{sec6}
In this paper, we computed a correlation measure called reflected entropy for an evaporating black hole. Unlike Page curve, as a measure of bipartite mixed state, reflected entropy can be computed between black hole and radiation, black hole and black hole, radiation and radiation. We have examined these curves in three different models: 3-side wormhole model, EOW brane model and JT gravity+CFT model.

For 3-side wormhole model, we calculated reflected entropy holographically with the wedge cross section and found that reflected entropy is dual to island cross sections, Reflected entropy $\sim$ Island cross section.
This provides a holographic dual of reflected entropy for quantum states living on multi-boundaries. On the other side, it implies that reflected entropy is associated to island cross section if an island can potentially appear in a gravitational system.
 By plotting the time dependence of reflected entropy, we found that correlation between (part of) radiation and black hole increases at early time and then decreases to zero, similar to the Page curve for entanglement entropy. It is actually bounded by the Page curve and encounters a transition after Page time. We also found that reflected entropy between radiation and radiation, jumps from zero to a finite value at Page time and increases until the black hole evaporates out.~\footnote{It means that the black hole disappears and the whole radiation becomes pure.}

For the EOW brane model, we found similar behaviors of reflected entropy curves as in 3-side wormhole model. Particularly, the reflected entropy between radiation and radiation increases at early time, jumps up at Page time and saturates at late time. The reflected entropy between radiation and black hole also encounters a transition after Page time.

We proposed a quantum extremal cross section (QECS) formula for the exact reflected entropy in AdS/CFT, as an analog of QES formula. To compute reflected entropy in more general gravitational systems, we employ the island formula of Von Neumann entropy in eternal black hole plus CFT model and obtain a generalized formula for reflected entropy with island cross section as its area term. Interestingly, the reflected entropy curve between the left-side black hole and the left-side radiation is nothing but Page curve. We also found that the reflected entropy between the left-side black hole and the right-side black hole decreases during evaporation until vanishing. And similar to the EOW-brane case, the reflected entropy between radiation and radiation increases at early time and saturates at late time, though we drew this result from the bounds of reflected entropy instead of calculating it explicitly.

Several future questions are in order: First, a pure CFT justification of our results in 3-side wormhole model, which will confirm our holographic computation and also help to reconstruct island cross section from CFT. Second, generalize our results to multipartite case using similar ideas in~\cite{Umemoto:2018jpc,Chu:2019etd}. This will involve a generalized correlation measure for the multi-boundary wormhole states and also a multipartite generalization of the QECS formula and its island version. Third, extent our results to other asymptotic flat models and higher-dimensions where we expect to find similar reflected entropy curves for an evaporating black hole. We hope to report the progress in future publications.
\section*{Acknowledgements}
We are grateful for useful discussion with our group members in Fudan University. This work is supported by NSFC grant 11905033.
\appendix
\section{The length formulas of $L_1$ and $L_2$}\label{L1-L2-formulas}
\par
The length formula of the geodesic $L_1$ can be worked out in the covering space Fig.\ref{covering2}. Since we identify $g_a\sim g_b$, the two brown curves of $L_1$ in Fig.\ref{covering2} are joint smoothly. Note that the complex coordinates of two endpoints of $L_1$ are $is_1$ and $is_2$. To find the formula of the geodesic that intersects these two points, one can map $is_2$ to the outside of the fundamental region using the generator which identifys $g_a$ and $g_b$ (\ref{gamma2}). It can be written in a simple form as
\begin{equation}
z \rightarrow \frac{D_aD_b}{X_a-z}+X_b.
\end{equation}
\par
Therefore, we have to find the geodesic which intersects both $is_1$ and $\frac{D_aD_b}{X_a-is_2}+X_b$. Since any geodesics in covering space are parts of semicircles with a center on the horizontal axis, we can easily determine the location of the center, which is given by
\begin{equation}
X_C = \frac{D_a^2 D_b^2 + 2 D_a D_b X_a X_b - (s_1^2 - X_b^2)(s_2^2 + X_a^2)}{2\left[D_a D_b X_a + X_b(s_2^2+X_a^2)\right]}.
\end{equation}
Its radius is obtained as the distance from the center to one endpoint
\begin{equation}
D_C = \sqrt{X_C^2 + s_1^2}.
\end{equation}
Next we have to compute the angles of the two intersection points. Then with the formula $\int_{\theta_1}^{\theta_2}d\theta / \sin{\theta}$ to compute the length of a portion of the semicircle, one can get the length formula of $L_1$
\begin{equation}
L_1(s_1,s_2) = \log \tan \left[ \frac{\pi}{2} - \frac{1}{2} \arcsin \frac{s_1}{D_C}\right] - \log \tan \left[\frac{1}{2} \arcsin \frac{s_2 D_a D_b}{D_C(s_2^2+X_a^2)}\right].
\end{equation}
It is worth noting that if we use the length formula in a general context which contains geometries with different AdS radiuses, we must multiply the result by the AdS radius of the geometry to which the curve belongs.

	\begin{figure}[htbp]
	\centering
	\includegraphics[scale=0.4]{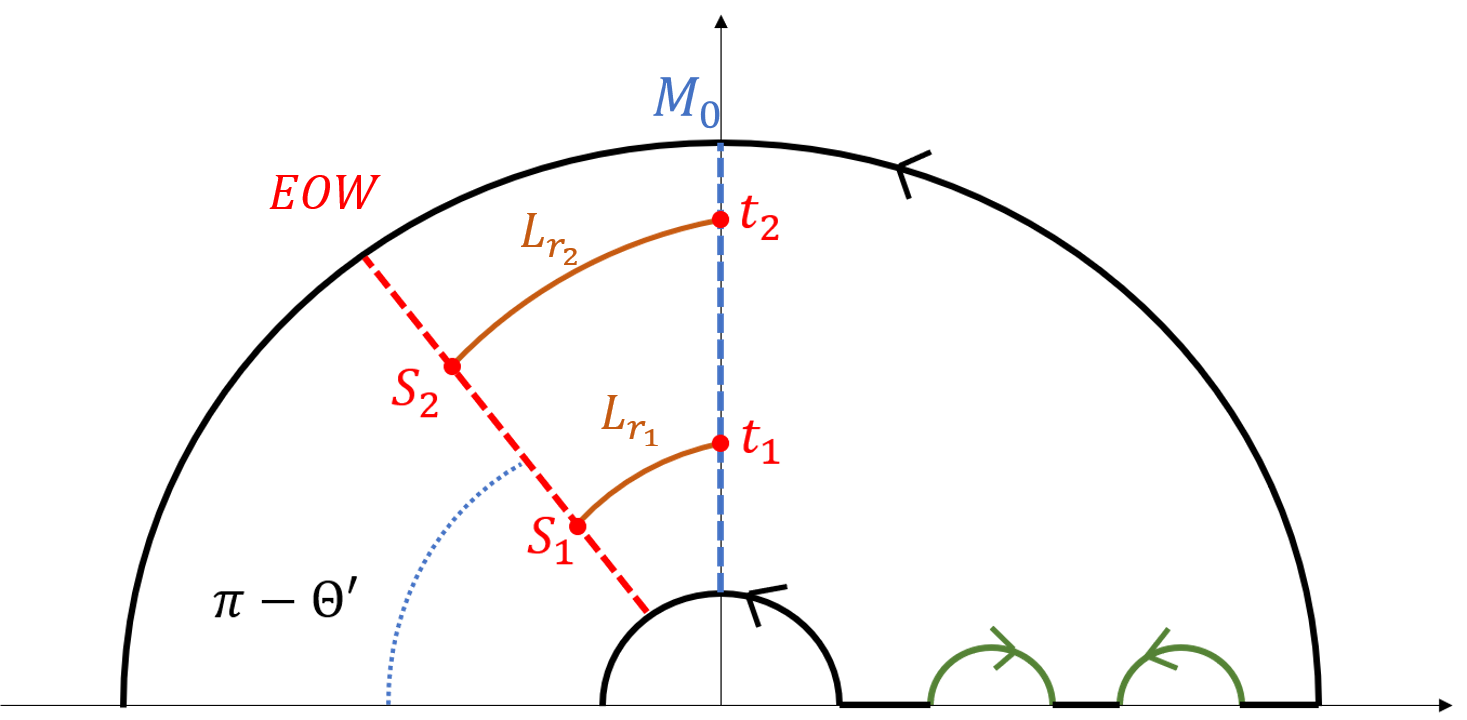}\\
	\caption{The covering space construction of the real geometry part (white region) in Fig.\ref{eowRRphase2Wormhole} is depicted here. The length of the blue vertical line matches that of $M_0$ in Fig.\ref{eowRRphase2Wormhole}.}
	\label{covering5}
	\end{figure}

\par
The geodesic $L_2$ in Fig.\ref{eowRRphase2Wormhole} consists of two parts, one part $L_{in}$ in the inception geometry (blue shaded region) and the other part $L_{re}$ in the real geometry (white region). They have two points of intersection on the EOW brane (red dashed circle). Let the intersection points be $s_1$ and $s_2$, the length of $L_{in}$ can be worked out using the same method as above, which is given by~\cite{Balasubramanian:2020hfs}
\begin{equation}
L_{in}(s_1,s_2)=\log \tan \left[ \frac{\pi}{2} - \frac{1}{2}\arcsin \frac{s_1 \sin\Theta}{D_I}\right]  -  \log \tan \left[  \frac{1}{2}\arcsin\frac{s_2 D_a D_b \sin{\Theta}}{D_I(s_2^2-2s_2 X_a \cos{\Theta}+X_a^2)}  \right],
\end{equation}
where $\Theta$ is the angle of the EOW brane in Fig.\ref{inceptionCovering} and $D_I$ is the radius of the semicircle. It is given by
\begin{equation}
D_I^2 = (X_I - s_1 \cos \Theta )^2 + s_1^2 \sin^2\Theta,
\end{equation}
where $X_I$ is the location of the center
\begin{equation}
\footnotesize
    X_I = \frac{D_a^2 D_b^2+2 s_2 \cos\Theta \left(s_1^2
   X_a-X_b (D_a D_b+X_a X_b)\right)+2 D_a
   D_b X_a X_b-(s_1-X_b) (s_1+X_b)
   \left(s_2^2+X_a^2\right)}{2 \left(-\cos\Theta \left(D_a
   D_b s_2+s_1 \left(s_2^2+X_a^2\right)+2 s_2
   X_a X_b\right)+D_a D_b X_a+s_1 s_2
   X_a \cos 2\Theta+s_1 s_2 X_a+X_b
   \left(s_2^2+X_a^2\right)\right)} .
\end{equation}
\par
$L_{re}$ consists of two geodesics which are depicted as brown curves in the covering space of real geometry (Fig.\ref{covering5}). The two endpoints of one of these curve determine the semicircle it belongs to. Take the lower curve $L_{r_1}$ for example, the center determined by the endpoints $s_1\exp(i\Theta')$ and $it_1$ is given by
\begin{equation}
X_{r_1} = \frac{s_1^2-t_1^2 }{2 s_1\cos\Theta'},
\end{equation}
where $\Theta'$ is the angle of the EOW brane (red dashed line) in Fig.\ref{covering5}, which is related to the location of the EOW brane in the real geometry $r_t$ by
\begin{equation}
2\pi r_t=\frac{m_0}{\sin \Theta'}.
\end{equation}
And the radius is given by
\begin{equation}
D_{r_1}=\sqrt{X_{r_1}^2 + t_1^2}.
\end{equation}
The length formula of the lower curve can then be worked out as
\begin{equation} \label{Lr-formula}
L_{r_1}=\log \tan \left[ \frac{\pi}{2} - \frac{1}{2}\arcsin \frac{s_1\sin\Theta'}{D_{r_1}}\right] - \log \tan \left[ \frac{1}{2} \arcsin \frac{t_1}{D_{r_1}}   \right].
\end{equation}
Note that the length formula of the upper curve $L_{r_2}$ takes the same form as (\ref{Lr-formula}) with $s_1$ and $t_1$ replaced by $s_2$ and $t_2$ respectively. Therefore, the entropy of $L_2$ is given by
\begin{equation}
S_{L_2}(s_1,s_2,t_1,t_2) = \frac{l' L_{in}}{4G_N'} + \frac{l(L_{r_1}+L_{r_2})}{4G_N},
\end{equation}
where $l$ and $l'$ are AdS radiuses for the real and inception geometry respectively. The minimal entropy is obtained by minimize $L_2$ over $s_1$, $s_2$, $t_1$ and $t_2$.
\section{The length formula of $L_I$}\label{LI-section}
The length of geodesic $L_I$ can be computed in the covering space Fig.\ref{covering3}. Again, the two brown curves are joint smoothly since we identify $g_1$ and $g_2$ using the generator $\gamma_1(z)=\frac{D_2}{D_1}z=\mu^2 z$. The two end points $s_1, s_2$ of $L_I$ are freely located on the horizon $M_1$ (blue dashed curve), which equation is given by~\cite{Balasubramanian:2020hfs}
\begin{equation}
M_1(\lambda)=\frac{X_a+X_b}{2} + \frac{1}{2} \exp(i\lambda)\sqrt{(X_a-X_b)^2-4D_aD_b}.
\end{equation}
Let the coordinates of $s_1$ and $s_2$ be $(x_1, y_1)$ and $(x_2,y_2)$ respectively. One can use $\gamma_1^{-1}$ to map $s_2$ to the point $s_2'$ with coordinate $(x_2/\mu^2,y_2/\mu^2)$. Then the geodesic $L_I$, which is a portion of a semicircle with a center on the horizontal axis again, is determined by the two points $s_1$ and $s_2'$. The center and radius of the semicircle is given by
\begin{equation}
X_I=\frac{D_a(D_b - D_b \mu^4)-x_2 X_a-x_2 X_b + X_a X_b + \mu^4 \left[x_1(X_a+X_b)-X_a X_b\right]}{2\mu^4 x_1 - 2\mu^2 x_2}
\end{equation}
\begin{equation}
D_I=\sqrt{(X_I-x_1)^2+y_1^2}
\end{equation}
Next we compute the angles of the two endpoints $s_1$ and $s_2'$ and then use them to get the length formula of $L_I$
\begin{equation}\label{LI-formula}
L_I(s_1, s_2)=\log \tan \left[\frac{\pi}{2} - \frac{1}{2} \arcsin(\frac{y_2}{\mu^2 D_I})\right] - \log \tan \left[\frac{1}{2}\arcsin(\frac{y_1}{D_I})\right].
\end{equation}
\section{The length formula of $L_p$}\label{Lp-formula}
\begin{figure}[htbp]
	\centering
	\includegraphics[scale=0.4]{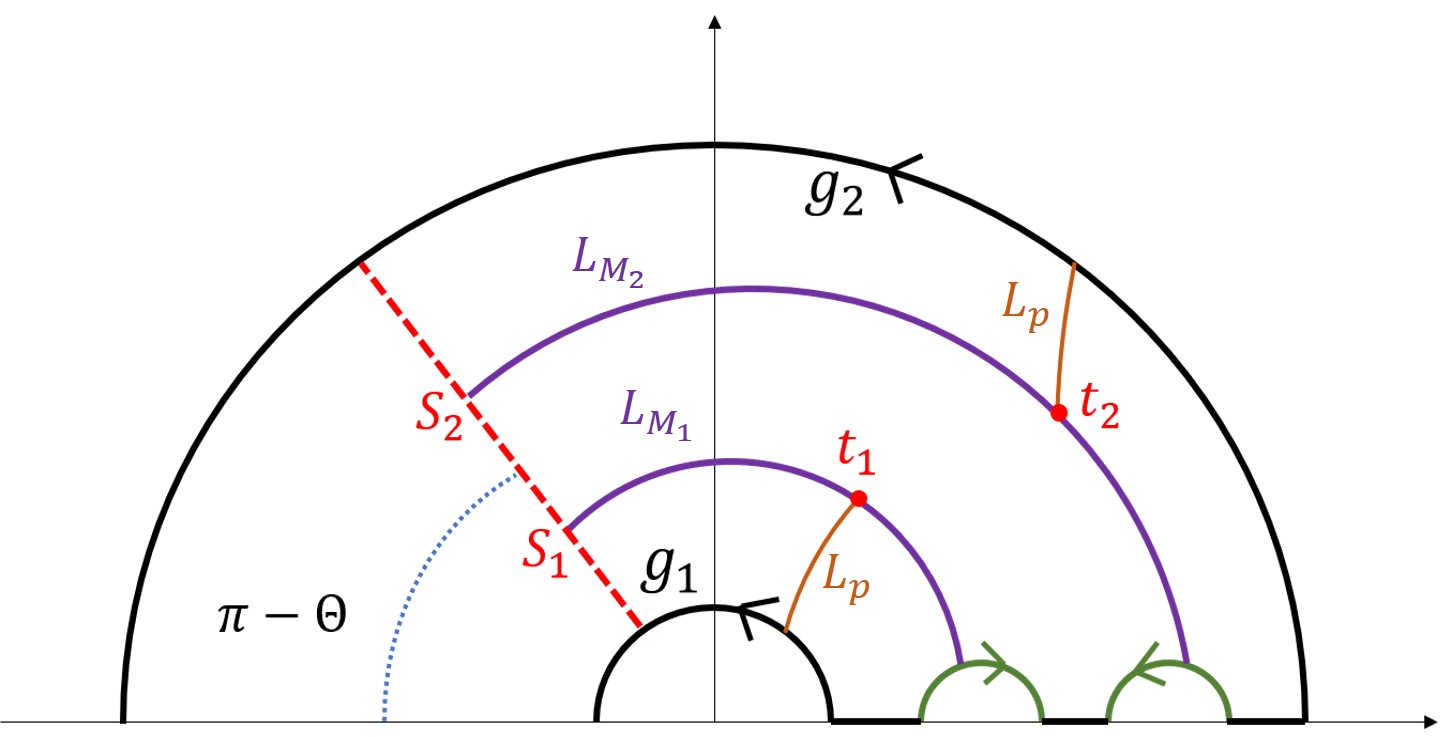}\\
	\caption{$L_{M_1}$ in Fig.\ref{eowRBphase2} is depicted as two purple arcs in the covering space. The geodesic $L_p$ is represented by the brown curves with two intersection points on $L_{M_1}$.}
	\label{covering4}
	\end{figure}

The geodesic $L_p$ in Fig.\ref{eowRBphase2} has two end points $t_1$ and $t_2$ on the purple curves $L_{M_1}$ which is depicted in the covering space Fig.\ref{covering4}. The brown curves are connected smoothly through the generator $\gamma_1$ that identifys $g_1$ and $g_2$. To get the length formula of $L_p$ we first have to figure out the equation of the purple curves which are parts of two semicircles with centers on the horizontal axis. Using the method in previous sections we can see that the lower purple curve is determined by the two points $s_1\exp(i\Theta)$ and $\frac{D_aD_b}{X_a-s_2\exp(i\Theta)}+X_b$. The higher purple curve is determined by the two points $s_2\exp(i\Theta)$ and $\frac{D_a D_b}{X_b-s_2\exp(i\Theta)}+X_a$. The centers and radiuses can then be easily obtained so we can get the equations of the two curves.
\par
Let the coordinates of $t_1$ and $t_2$ be $(x_1, y_1)$ and $(x_2, y_2)$ respectively. We can use $\gamma_1^{-1}$ to map $t_2$ to the point $t_2'$ with coordinate $(x_2/\mu^2,y_2/\mu^2)$. This is the same case as in Appendix \ref{LI-section}, thus the length formula of $L_p$ takes the same form as (\ref{LI-formula}).
\section{Bounds on reflected entropy}\label{bounds}

	\begin{figure}[htbp]
	\centering
	\includegraphics[scale=0.4]{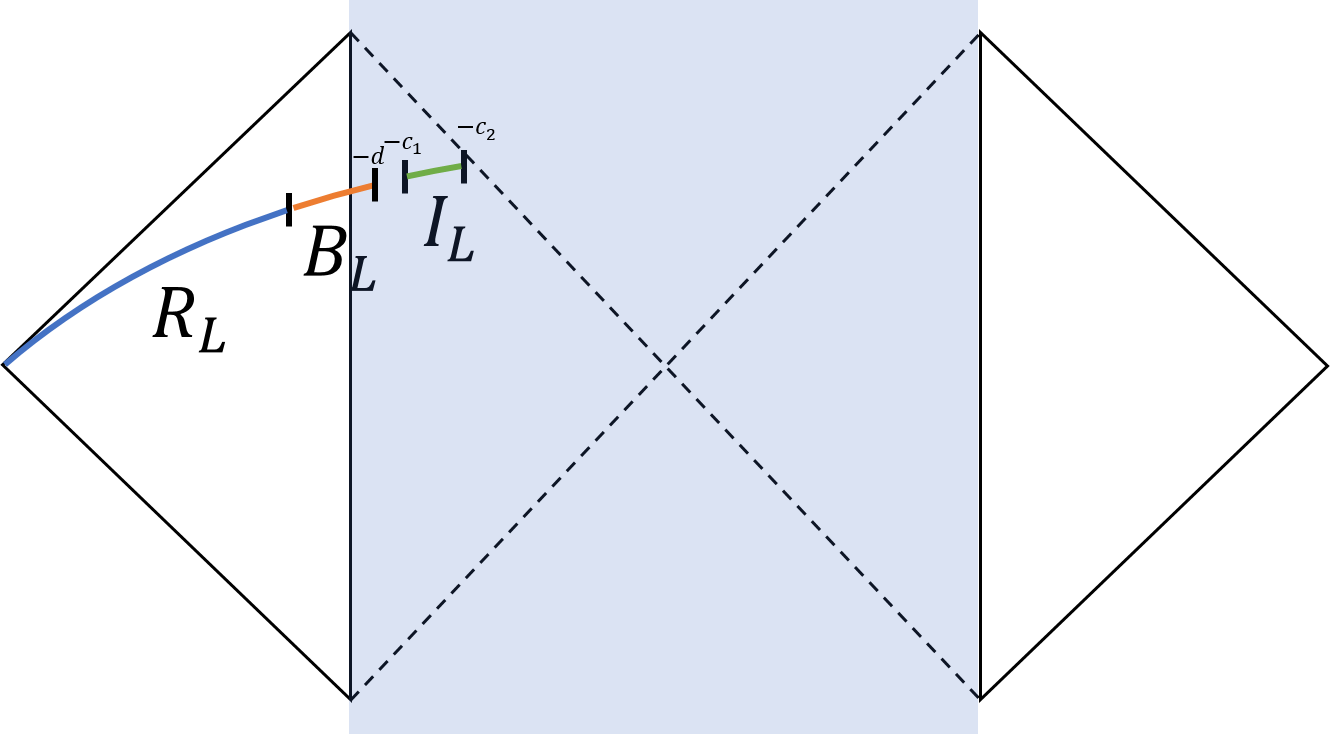}\\
	\caption{Radiation and black hole on one side. The point $(-d,-t+\pi i)$ is the quantum extremal surface of the black hole. The island of radiation is the interval bounded by the points $(-c_1,-t+\pi i)$ and $(-c_2,-t+\pi i)$.}
	\label{jtAC2}
	\end{figure}

We would like to compare $S_R(R_L:B_L)$ with its upper bound $2\min \{ S(R_L),S(B_L) \}$ and the lower bound $I(R_L:B_L)$. Since we only consider one side of the eternal black hole, the bounds should be independent of time. With variables shown in Fig.\ref{jtAC2}, the entropy of the left black hole is 
\begin{equation}
\label{ebl}
S(B_L)=S_0+\frac{\phi_r}{\tanh d}+\frac{c}{6}\ln \frac{4\sinh^2 \frac{b+d}{2}}{\sinh d}-\frac{c}{6}\ln \epsilon_{UV}\ ,
\end{equation}
where the quantum extremal surface $(-d,-t+\pi i)$ is determined by the QES condition
\begin{equation}
\label{QESd}
\sinh d=\frac{6\phi_r}{c}\frac{\sinh \frac{a+d}{2}}{\sinh \frac{a-d}{2}} \ .
\end{equation}
To evaluate the entropy of $R_L$, we first make such a cut-off that the left boundary of $R_L$ is $(\Lambda ,-t+\pi i)$, with $\Lambda$ sufficiently large. Then $S(R_L)$ is given by
\begin{equation}
S^{\text{island}}(R_L)=2S_0+\frac{\phi_r}{\tanh c_2}+\frac{\phi_r}{\tanh c_1}+\frac{c}{6}\ln \frac{4(e^{-c_1}-e^{-c_2})^2(e^b-e^{-c_1})^2}{(1-e^{-2c_2})e^b(e^b-e^{-c_2})^2(1-e^{-2c_1})}-\frac{c}{3}\ln \epsilon_{UV}+\frac{c}{6}\Lambda\ ,
\end{equation}
\begin{equation}
S^{\text{no island}}(R_L)=-\frac{c}{6}b-\frac{c}{3}\ln \epsilon_{UV}+\frac{c}{6}\Lambda \ ,
\end{equation}
\begin{equation}
S(R_L)=\min \{S^{\text{island}}(R_L),S^{\text{no island}}(R_L)\} \ .
\end{equation}
Note that $|w_1|=e^{-c_1},|w_2|=e^{-c_2}$. So the QES equations determining $c_1$ and $c_2$ can be written in terms of $|w_1|$ and $|w_2|$ as follows :
\begin{equation}
\label{QESc1}
\begin{split}
-|w_1| \partial_{|w_1|} S^{\text{island}}(R_L)=&-|w_1|(\frac{4\phi_r |w_1|}{(1-|w_1|^2)^2}+\frac{c}{3}\frac{1}{|w_1|-|w_2|}+\frac{c}{3}\frac{1}{|w_1|-e^{b}}-\frac{c}{3}\frac{|w_1|}{|w_1|^2-1})\\
=&0 \ ,
\end{split}
\end{equation}
\begin{equation}
\label{QESc2}
\begin{split}
-|w_2| \partial_{|w_2|} S^{\text{island}}(R_L)=&-|w_2|(\frac{4\phi_r |w_2|}{(1-|w_2|^2)^2}+\frac{c}{3}\frac{1}{|w_2|-|w_1|}-\frac{c}{3}\frac{1}{|w_2|-e^{b}}-\frac{c}{3}\frac{|w_2|}{|w_2|^2-1})\\
=&0 \ .
\end{split}
\end{equation}
Compared with $S(B_L)$, $S(R_L)$ has an IR divergent term $\frac{c}{6}\Lambda$, so we conclude that $S(B_L)$ is less than $S(R_L)$. Therefore, $2\min \{ S(R_L),S(B_L) \}=2S(B_L)$ which is the upper bound of $S_R(R_L:B_L)$. 

For the lower bound $I(R_L:B_L)=S(B_L)+S(R_L)-S(R_L\cup B_L)$, we should also calculate $S(R_L\cup B_L)$. The interval of $R_L\cup B_L$ could be assumed as $ (\infty_L,P_0]$ with $P_0=(-f,-t+\pi i)$ to be extremized. The entropy of this interval is
\begin{equation}
\label{sac}
S(R_L\cup B_L)=S_0+\frac{\phi_r}{\tanh f}+\frac{c}{6}\ln \frac{2}{1-e^{-2f}}-\frac{c}{6}\ln \epsilon_{UV}+\frac{c}{6}\Lambda \ .
\end{equation}
By extremizing (\ref{sac}), we get the condition of $f$ (in terms of $|w_0|=e^{-f}$):
\begin{equation}
\label{QESf}
\begin{split}
-|w_0|(\frac{4\phi_r |w_0|}{(1-|w_0|^2)^2}+\frac{c}{3}\frac{|w_0|}{1-|w_0|^2})=0\ ,
\end{split}
\end{equation}
which has only one solution $|w_0|=0$ because $|w_0|=e^{-f}<1$ which leads to positivity of the bracket term in (\ref{QESf}). In other words, $R_L\cup B_L$ is just the left half line of the time slice. So 
\begin{equation}
S(R_L\cup B_L)=S_0+\phi_r+\frac{c}{6}\ln 2-\frac{c}{6}\ln \epsilon_{UV}+\frac{c}{6}\Lambda \ .
\end{equation}

\end{document}